\documentclass[%
aps,
prb,%
reprint,%
amsmath,amssymb,%
floatfix,%
nofootinbib,%
superscriptaddress%
]{revtex4-2}

\usepackage[T1]{fontenc}
\usepackage[utf8]{inputenc}
\usepackage[english]{babel}
\usepackage{mathrsfs}
\usepackage{bbm}
\usepackage{graphicx}
\usepackage{xcolor}
\usepackage{physics}
\usepackage{bm}
\usepackage{enumerate}

\usepackage[colorlinks,
            linkcolor=blue,    
            citecolor=blue,  
            urlcolor=blue,
 	        bookmarks=true,        
	        bookmarksopen=true,    
	        bookmarksnumbered=true,
]{hyperref}

\renewcommand*\d{\mathop{}\!\mathrm{d}}
\newcommand{\dirac}[1]{\,\delta\!\left(#1\right)}

\newcommand{\av}[1]{\left\langle#1\right\rangle}

\newcommand{\conj}[1]{\overline{#1}}

\newcommand{\mypath}{Figs}

\begin{document}

\title{Spectral transitions and universal steady states in random Kraus maps and circuits}

\author{Lucas  S\'a}
\email{lucas.seara.sa@tecnico.ulisboa.pt}

\affiliation{CeFEMA, Instituto Superior T\'ecnico, Universidade de Lisboa, Av.\ Rovisco Pais, 1049-001 Lisboa, Portugal}

\author{Pedro Ribeiro}
\email{ribeiro.pedro@tecnico.ulisboa.pt}

\affiliation{CeFEMA, Instituto Superior T\'ecnico, Universidade de Lisboa, Av.\ Rovisco Pais, 1049-001 Lisboa, Portugal}
\affiliation{Beijing Computational Science Research Center, Beijing 100193, China}

\author{Tankut Can}
\email{tankut.can@gmail.com}

\affiliation{Initiative for the Theoretical Sciences, The Graduate Center, CUNY, New York, NY 10016, USA}

\author{Toma\v z Prosen}
\email{tomaz.prosen@fmf.uni-lj.si}

\affiliation{Department of Physics, Faculty of Mathematics and Physics, University of Ljubljana, Ljubljana, Slovenia}
    
\begin{abstract}
The study of dissipation and decoherence in generic open quantum systems recently led to the investigation of spectral and steady-state properties of random Lindbladian dynamics. A natural question is then how realistic and universal those properties are. Here, we address these issues by considering a different description of dissipative quantum systems, namely, the discrete-time Kraus map representation of completely positive quantum dynamics.
Through random matrix theory (RMT) techniques and numerical exact diagonalization, we study random Kraus maps, allowing for a varying dissipation strength, and their local circuit counterpart. 
We find the spectrum of the random Kraus map to be either an annulus or a disk inside the unit circle in the complex plane, with a transition between the two cases taking place at a critical value of dissipation strength.
The eigenvalue distribution and the spectral transition are well described by a simplified RMT model that we can solve exactly in the thermodynamic limit, by means of non-Hermitian RMT and quaternionic free probability. The steady state, on the contrary, is not affected by the spectral transition. It has, however, a perturbative crossover regime at small dissipation, inside which the steady state is characterized by uncorrelated eigenvalues. At large dissipation (or for any dissipation for a large-enough system), the steady state is well described by a random Wishart matrix. The steady-state properties thus coincide with those already observed for random Lindbladian dynamics, indicating their universality.
Quite remarkably, the statistical properties of the local Kraus circuit are qualitatively the same as those of the nonlocal Kraus map, indicating that the latter, which is more tractable, already captures the realistic and universal physical properties of generic open quantum systems.
\end{abstract}

\maketitle

\section{Introduction}
The controlled manipulation of a large number of quantum degrees of freedom is becoming an experimental reality, driven by quantum information processing and quantum sensing applications. 
However, statistical deviations to controlling protocols and interactions with the environment are, to some degree, unavoidable. 
The ensuing relaxation and decoherence effects are responsible for computation errors and limit the accuracy of sensing devices. 
Such processes are modeled by imperfect quantum channels, which generalize unitary operations to an open-system setup. 
Under rather general assumptions, these completely positive maps can be considered memoryless, thus acquiring the general form of so-called Kraus maps~\cite{kraus1983}.
However, in most cases, the exact non-unitary dynamical map is not known, preventing concrete modeling of these effects.

For closed quantum systems, a fruitful approach, whenever the microscopic theory is unknown, has been to rely on universal properties dictated solely by the systems' symmetry and by their chaotic or integrable nature. 
The agreement of quantities such as level spacing statistics is so striking that the conformity of generic (i.e., non-integrable) quantum systems to random matrix theory (RMT) has been formulated in the form of the quantum chaos conjecture~\cite{bohigas1984}, whereas regular (integrable) dynamics systems are expected to follow a Poissonian distribution~\cite{berry1977}. 

While the quantum chaos conjecture has been extensively used to model closed Hamiltonian systems (for a comprehensive review see Ref.~\cite{guhr1998} and references therein), a systematic application of RMT to the non-Hermitian generators of open quantum systems is more recent and still under development. (We note, however, that RMT approaches to open \emph{scattering} systems have a long and fruitful history; for reviews, see Refs.~\cite{beenaker1997,beenaker2015,schomerus2016}.)
Following pioneering works on non-Hermitian Hamiltonians~\cite{sokolov1988,sokolov1989,haake1992,lehmann1995}, some studies focused on structureless discrete-time quantum maps~\cite{bruzda2009,bruzda2010}, and decoherence effects induced by a random environment~\cite{gorin2003,gorin2008,xu2019}. 
Recently, the focus has shifted to continuous-time random Lindblad dynamics of Markovian open quantum systems~\cite{denisov2018,can2019prl,can2019jphysa,sa2020RL,wang2020}. 
These efforts led to a thorough numerical characterization of the spectral support, the spectral density, spectral gaps, and steady states, complemented by analytical calculations based on perturbative arguments and non-Hermitian RMT.

Given the wealth of recent results, two important overarching questions arise: 
\begin{enumerate}[(i)]
    \item How well do completely random models of open quantum systems describe real physical systems?
    \item How universal are the results found for the particular models studied so far?
\end{enumerate}
A step towards answering question (i) has been taken in Ref.~\cite{wang2020} by including a notion of locality into the random Lindbladian model. 
Notwithstanding, the study of chaotic discrete-time quantum maps has, up to now, been restricted to spatially structureless cases with no adjustable parameters controlling dissipation strength.
Regarding question (ii), a new universal classification measure for non-Hermitian RMT was put forward in Ref.~\cite{sa2019CSR}. It was shown to apply to random matrices as well as to realistic model examples, depending only on their symmetry --- a result that provided some of the first elements to answer question (i) (see also Refs.~\cite{akemann2019,hamazaki2019}).  
Furthermore, the independence of the shape of the complex Lindbladian spectral support from the particular sampling scheme was discussed in Ref.~\cite{denisov2018}. 
Meanwhile, other important questions remain unaddressed. One of the most pressing is whether the universal nature of the steady-state properties identified in Ref.~\cite{sa2020RL} is independent of the RMT ensemble or even from the continuous- or discrete-time description of the dynamics.

The goal of this paper is to tackle the aforementioned issues and make additional steps towards answering questions (i) and (ii). 
To this end, we consider a tractable statistical model featuring periodic discrete-time dynamics generated by a completely positive map as a model for generic Floquet dynamics in an open quantum system. 
We study the spectral and steady-state properties of structureless random Kraus maps and their local circuit counterparts, built from matrices of the circular unitary ensemble (CUE)~\cite{dyson1962i,haake2013}. Our model is a more structured refinement of that in Ref.~\cite{bruzda2009}, allowing for a varying dissipation strength and to enforce locality onto the evolution. It also enables direct comparison with previously studied random Lindblad dynamics through the exponential map. 

Remarkably, the spectral density (one-point function) of the random quantum map can be analytically characterized, including the spectral boundaries (as done for random Lindblad dynamics~\cite{denisov2018}) and the spectral density (which was not yet accomplished in the Lindblad case). 
We find the complex-valued spectrum to assume either an annulus- or a disk-shaped distribution inside the unit circle. 
Varying the strength of dissipation leads to a transition from an annular to a disk spectral support. 
While different from the spectrum of the random Lindblad Liouvillian (which assumes a lemon-like shape), the disk/annular support we find here appears to be quite ubiquitous for complex discrete-time quantum maps. 
In contrast, the steady-state properties are essentially the same as in the random Lindblad case, suggesting a high degree of universality. 
The steady state is not affected by the spectral transition, displaying, however, the same perturbative-to-RMT crossover regime at small dissipation, already observed for Lindbladian dynamics~\cite{sa2020RL}. Finally, considering a 1D quantum dissipative circuit with only local interactions 
does not qualitatively change the results found in the unstructured case, pointing to a common universality class for the statistical properties of spatially unstructured random Kraus maps and local circuits.

The paper is organized as follows. In Sec.~\ref{sec:model} we define the model studied. Nonlocal 0D Kraus maps and local 1D Kraus circuits are then separately analyzed in detail in Secs.~\ref{sec:0d} and \ref{sec:1d}, respectively, before we draw our conclusions in Sec.~\ref{sec:conclusion}. The Appendix presents a derivation of the spectral support and eigenvalue distribution of the RMT model used to analytically describe the 0D Kraus map.

\begin{figure*}[tp]
	\centering
	\includegraphics[width=0.99\textwidth]{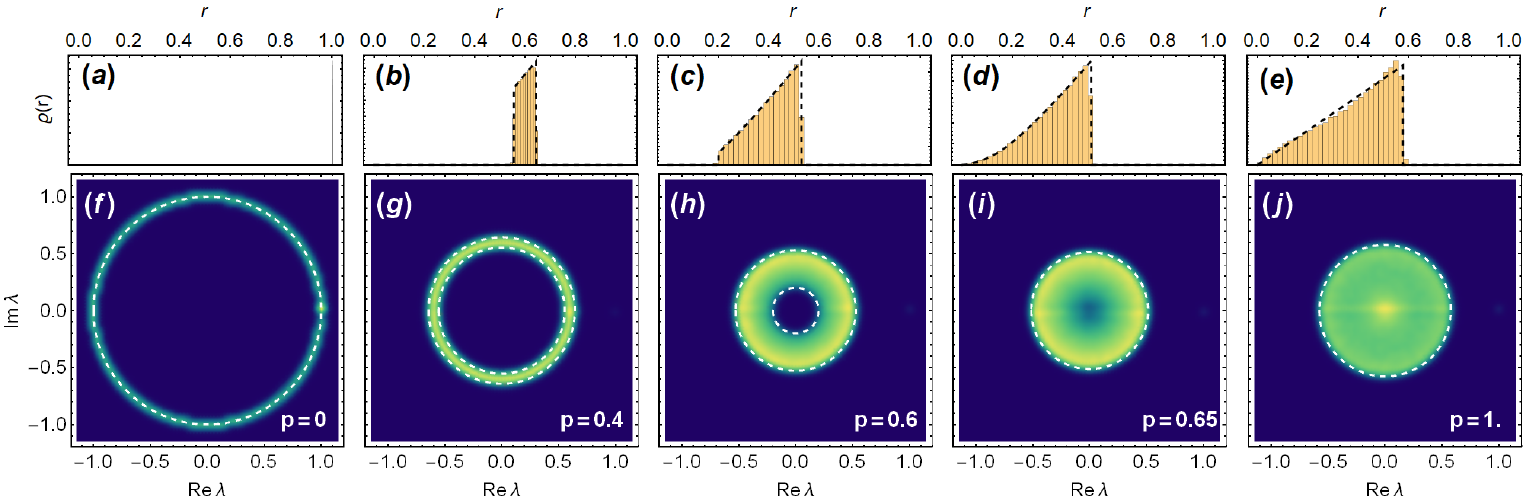}
	\caption{Global spectrum of the fully-connected quantum map, for different values of $p\in[0,1]$, $N=50$, and $d=3$. The eigenvalues are obtained from exact diagonalization of the map (\ref{eq:0d_map}). Ensemble averaging is performed such as to always obtain at least $10^5$ eigenvalues. (a)--(e): radial eigenvalue distribution (density); the dashed black line is given by the theoretical prediction of Eq.~(\ref{eq:pKraus_rhoR_RMT}). (f)--(j): eigenvalue density in the complex plane; the outer (inner) dashed line depicts the outer (inner) circular boundary of radius $R_+$ ($R_-$), given by Eq.~(\ref{eq:pKraus_rpm_analytic}).}
	\label{fig:pKraus_global_spectrum}
\end{figure*}

\section{Parametric Random Kraus Operators}
\label{sec:model}

Consider a quantum system in some initial state $\rho_0$. The action of a quantum dynamical map $\Phi$ (with $k$ decay channels) leads to a new state $\rho_1=\Phi\left(\rho_0\right)=\sum_{\mu=1}^kK_\mu\rho_0 K_\mu^\dagger$, where the Kraus operators $K_\mu$ are subjected to the trace-preservation constraint $\sum_{\mu=1}^kK_\mu ^\dagger K_\mu=\mathbbm{1}$~\cite{nielsen2002,bengtsson2017}. The successive action of the quantum map, $t$ times, leads to the final state $\rho_t=\Phi\left(\rho_{t-1}\right)=\Phi^t\left(\rho_0\right)$. The superoperator $\Phi$ admits the matrix representation
\begin{equation}\label{eq:def_map}
\Phi=\sum_{\mu=1}^k K_\mu\otimes K_\mu^*.
\end{equation}

We parametrize the deviation of the Kraus map from unitarity through a parameter $p\in[0,1]$, such that $p=0$ corresponds to unitary evolution, while $p=1$ corresponds to the case of a structureless quantum map studied in Refs.~\cite{bruzda2009,bruzda2010}. To this end, we consider two types of Kraus operators $K_\mu$ (in total $k=1+d$):
\begin{equation}\label{eq:Kraus_def}
\begin{split}
&K_0=\sqrt{1-p}\,U,\qquad \text{with}\quad  U^\dagger U=\mathbbm{1},\\
&K_j=\sqrt{p}\,M_j,\qquad \text{with}\quad \sum_{j=1}^d M_j^\dagger M_j=\mathbbm{1}.
\end{split}
\end{equation}
Here, $U$ is an $N\times N$ unitary matrix, while the $d$ Kraus operators $M_j$ are constructed as truncations of enlarged $Nd\times Nd$ unitary matrices~\cite{zyczkowski2000}, following Ref.~\cite{bruzda2009}:  generate an $Nd\times Nd$ random unitary matrix $V$, formed by $d^2$ blocks $V_{ij}$, $i,j=1,\dots,d$, of dimension $N\times N$,
\begin{equation}\label{eq:blocksU}
V=
\begin{pmatrix}
V_{11} & V_{12} & \cdots & V_{1d} \\
V_{21} & V_{22} & \cdots & V_{2d} \\
\vdots & \vdots & \ddots & \vdots \\
V_{d1} & V_{d2} & \cdots & V_{dd}
\end{pmatrix},
\end{equation}
and take the $d$ Kraus operators to be the blocks of the first (block-) column, i.e., $M_j=V_{j1}$. The constraint $\sum_{j=1}^dM_j ^\dagger M_j=\mathbbm{1}$ is automatically satisfied because of the orthonormality of the columns of $V$.\footnote{
	More generally, we could take the $j$th Kraus operator as any (normalized) linear combination of (block-) columns of $V$, i.e., $M_j=\sum_{\alpha=1}^d\psi_\alpha V_{j\alpha}$, with $\sum_{\alpha=1}^d\abs{\psi_\alpha}^2=1$. Without loss of generality, we set $\psi_1=1$ and $\psi_\alpha=0$ for $\alpha\neq1$ from now on.}
By construction, the Kraus operators $K_\mu$, $\mu=0,\dots,d$, satisfy $\sum_\mu K_\mu^\dagger K_\mu=\mathbbm{1}$. In what follows, to construct a random Kraus map, we draw both $U$ and $V$ from the CUE.

\section{0D Random Quantum Maps}
\label{sec:0d}
First, we consider the most general random quantum map, without imposing any spatial structure. The quantum system can be understood as $N$ sites on a fully connected graph (and hence interpreted either as a 0D or $\infty$D system). Quantum maps defined on a 1D lattice are addressed in Sec.~\ref{sec:1d}. The matrix representation of the 0D quantum dynamical map is given by
\begin{equation}\label{eq:0d_map}
\Phi=(1-p)\,U\otimes U^*+p\sum_{j=1}^dM_j\otimes M_j^*.
\end{equation}

\begin{figure}[tp]
	\centering
	\includegraphics[width=0.8\columnwidth]{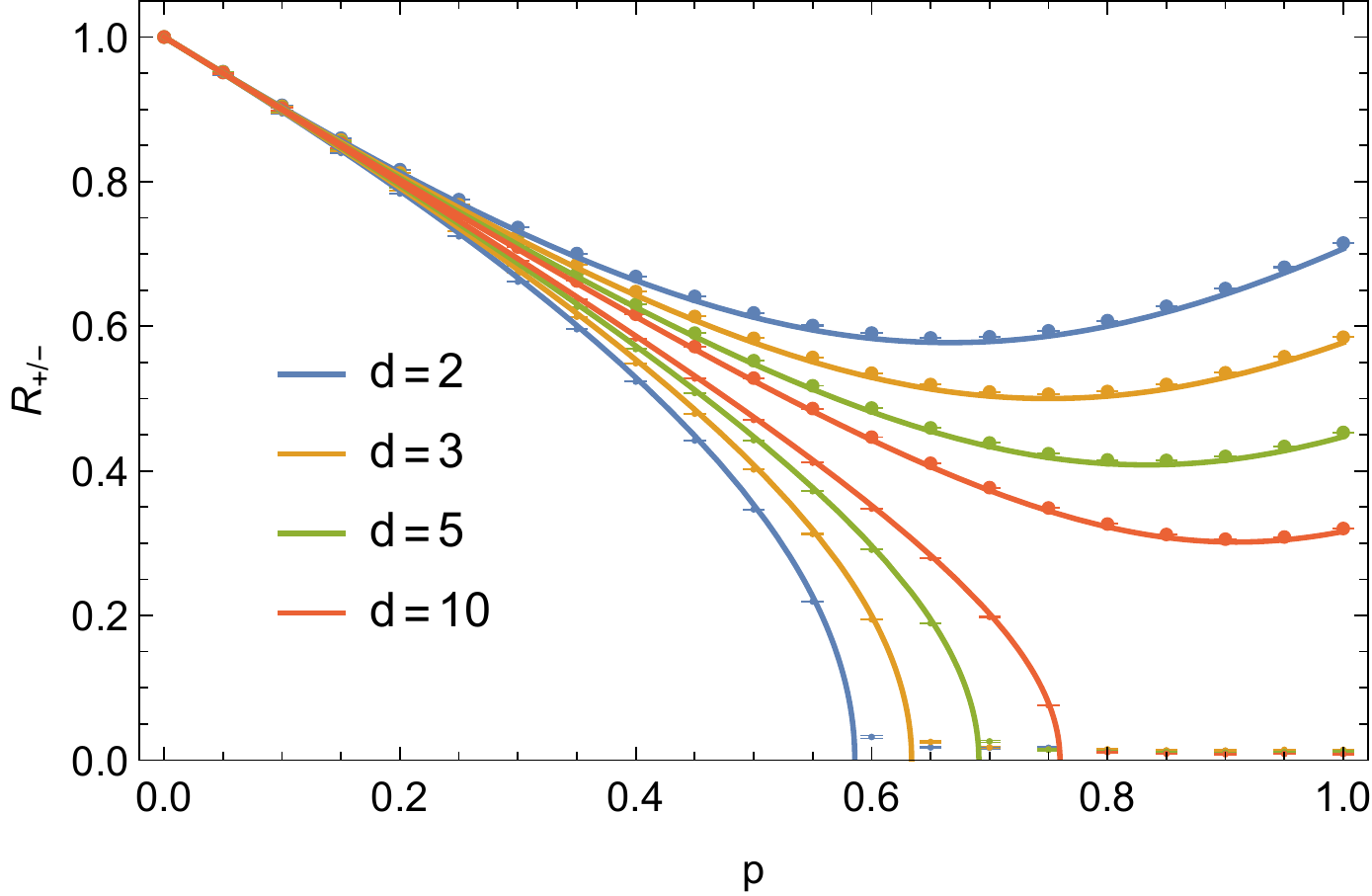}
	\caption{Inner and outer radius of the eigenvalue support of the fully-connected quantum map, as a function of $p$, for different $d$ and $N=60$. The points are the numerical results from exact diagonalization of $\Phi$ from Eq.~(\ref{eq:0d_map}), the solid lines give the analytical expressions from Eq.~(\ref{eq:pKraus_rpm_analytic}).}
	\label{fig:pKraus_rINOUT}
\end{figure}

\subsection{Spectral Properties}

\subsubsection{Numerical results}
The spectrum of $\Phi$ in the complex plane, obtained by exact diagonalization, is plotted in Fig.~\ref{fig:pKraus_global_spectrum}, for different values of $p$, $N=50$, and $d=3$. We note that the spectral distribution of the operator $\Phi$ is expected to be self-averaging when $N\to\infty$. We indeed observed that, for the system sizes considered, single realizations are already very close to the ensemble average that we plot in Fig.~\ref{fig:pKraus_global_spectrum}.

For $p=0$ all the spectral weight lies on the unit circle and for $p=1$ it uniformly covers the disk of radius $1/\sqrt{d}$. For intermediate values, it forms an annulus with well-defined inner and outer radius, $R_+$ and $R_-$, respectively. The annulus closes to a disk, i.e., $R_-=0$, at a finite value of $p_c<1$ depending only on $d$ and not on $N$, in the limit of large $N$. 
Using the ansatz explained in Sec.~\ref{subsec:RMT_model}, 
the value of $p_c$, and more generally the functions $R_\pm(p)$ can be computed analytically in the large-$N$ limit. 
Figure~\ref{fig:pKraus_rINOUT} shows the inner and outer radius of the eigenvalue distribution for several values of $d$ ($d=2$, $3$, $5$ and $10$).
With increasing $p$, the spectrum first contracts into a circle of radius smaller than $1/\sqrt{d}$, before spreading again and attaining a radius $1/\sqrt{d}$ at exactly $p=1$. 
The value of $p$ for which the minimum value is attained goes to $p=1$ as $d$ increases. 

Because of the rotational invariance of the spectrum, the spectral gap of $\Phi$, corresponding to the slowest decaying mode, is given by $\Delta = - \log\abs{R_+}$. 
The isolated eigenvalue at $\lambda = 1$ corresponds to the fixed point of $\Phi$, i.e., the steady state $\rho_{\text{SS}}$. This eigenvalue is not visible in the plots due to its vanishing spectral weight. 
For asymptotic long times, $\Delta$ is the rate at which a density matrix approaches the steady state $\rho_{\text{SS}}$ under repeated application of $\Phi$, i.e., $\rho_t - \rho_{\text{SS}} \propto e^{-\Delta t} $. 
Interestingly, $\Delta$ is a non-monotonous function of $p$, contradicting the naive expectation that a larger non-Hermitian component yields faster relaxation. 
The eigenmodes of $\Phi$ with the fastest decaying rates correspond to the smallest $\abs{\lambda}$. For $p>p_c$, a number of such modes instantaneously vanish after the first application of $\Phi$. On the contrary, for $p<p_c$ all modes have a finite lifetime.     

\subsubsection{RMT Model for the Quantum Map}
\label{subsec:RMT_model}

The form of Eq.~(\ref{eq:0d_map}) makes it difficult to directly determine  the spectral density of $\Phi$. 
Here, in order to analytically characterize the eigenvalue distribution of this quantum map, we analyze instead a tractable effective RMT model given by
\begin{equation}\label{eq:Phieff}
\tilde \Phi =(1-p)\,\mathbb{U}+\frac{p}{\sqrt{d}}\mathbb{G},
\end{equation}
where $\mathbb{U}$ is an $N^2\times N^2$ Haar-random unitary matrix and $\mathbb{G}$ is an $N^2\times N^2$ random matrix drawn from the Ginibre Unitary Ensemble (GinUE)~\cite{ginibre1965,haake2013} with unit variance.\footnote{
    The effective model~(\ref{eq:Phieff}) does not account for the eigenvalue $1$ (steady state), thus describing only the annular/disk-shaped bulk of the spectrum.
} 
The effective model~(\ref{eq:Phieff}) becomes exact in the double limit $N,d\to\infty$ and can be justified as follows. 

We start with the unitary contribution to Eqs.~(\ref{eq:0d_map}) and (\ref{eq:Phieff}). To analytically compute the spectral properties of $\tilde \Phi$, only two properties of $\mathbb{U}$ are used: it is unitary and it has a flat spectrum (see the Appendix for details of the computation). Therefore, for the purpose of our calculations, we can approximate $U\otimes U^*$ by $\mathbb{U}$ if the former also possesses these two properties. Since $U\otimes U^*$ is trivially unitary, it only remains to show that it has a flat spectrum, at least in the large-$N$ limit.
Now, $U\otimes U^*$ has eigenvalues $\exp{i\varphi_{\alpha\beta}}\equiv\exp{i(\theta_\alpha-\theta_\beta)}$, $\alpha,\beta=1,\dots,N$, where $\exp{i\theta_\alpha}$ are the eigenvalues of $U$, with flat spectral density $\varrho_U(\theta)=1/(2\pi)$ on the unit circle. In the large-$N$ limit, the spectral density of $U\otimes U^*$ is (we denote the spectral density of $U\otimes U^*$ by $\varrho_\otimes(\varphi)$ and the two-point function of $U$ by $R_2(\theta_1,\theta_2)$):
\begin{equation}
\begin{split}\label{eq:rho_otimes}
    \varrho_{\otimes}(\varphi)
    &=\int\d\theta_1\d\theta_2\, R_2(\theta_1,\theta_2) \dirac{\varphi-(\theta_1-\theta_2)}
    =2\pi R_2(\varphi)\\
    &=\frac{1}{2\pi}\frac{1}{1-1/N}\left(
    1-\frac{1}{N^2}\left(\frac{\sin(N\varphi/2)}{\sin(\varphi/2)}\right)^2
    \right),
\end{split}
\end{equation}
where we have used the translational invariance of the CUE two-point function, 
\begin{equation}
\begin{split}
    &R_2(\theta_1,\theta_2)
    =R_2(\theta_1-\theta_2)\\
    =&\frac{1}{N(N-1)}\left[
    \left(\frac{N}{2\pi}\right)^2-
    \left(\frac{1}{2\pi}\frac{\sin(N(\theta_1-\theta_2)/2)}{\sin((\theta_1-\theta_2)/2)}\right)^2
    \right].
\end{split}
\end{equation}
In the large-$N$ limit, the second term inside the brackets in Eq.~(\ref{eq:rho_otimes}) converges to $(2\pi/N)\dirac{\varphi}$, and hence, in this limit, $\varrho_\otimes(\varphi)\to1/(2\pi)$, as claimed.

Regarding the dissipative term in Eq.~(\ref{eq:Phieff}), it was conjectured in Ref.~\cite{bruzda2009} that it can be approximated, for large $N$ and large $d$, by $(1/\sqrt{d})\,\mathbb{G}$.\footnote{
	Our results show that $d=3$ can already be considered as the large-$d$ limit if $N$ is the order of a few tens.}
Alternatively, one can note that each $M_j$ is a truncation of a Haar-random unitary and hence its entries are independent and identically distributed Gaussian random variables with zero mean and variance $\sigma_M^2/N$, where $\sigma_M^2=1/(2d)$~\cite{zyczkowski2000}. ($M_j$ is supported on a disk of radius $1/\sqrt{d}$, but its eigenvalue density is not flat inside this disk; it is instead flat on the hyperbolic plane, whence there is an increase of the spectral density near the boundary of the disk.) Now, let $K_j=M_j\otimes M_j^*$. Then, the first moment of $K_j$ is 
\begin{equation}
    \mu^K_1=
    \frac{1}{N^2}\left\langle \Tr K_j \right\rangle =
    \frac{1}{N^2}\left\langle \abs{\Tr M_j}^2 \right\rangle =
    \frac{2\sigma_M^2}{N}=
    \frac{1}{N d},
\end{equation}
which is zero in the large-$N$ limit. The second moment (equivalently the second cumulant) is 
\begin{equation}
    \mu^K_2=
    \frac{1}{N^{2}}\left\langle \Tr K_j^2-\langle\Tr K_j\rangle^2\right\rangle =
    4\sigma_M^4=
    \frac{1}{d^2}
\end{equation}
plus corrections which vanish when $N\to\infty$. So, we arrive at the claim that $K_j$ is represented by a random matrix whose entries are random variables with zero mean and variance $4\sigma_M^4/N^2$. By the central limit theorem of non-Hermitian matrices~\cite{nica2006,mingo2017}, taking the sum of $d$ such matrices (which are almost independent since the unitary constraints become less relevant as the dimensions of the matrices grow) results in a $N^2\times N^2$ (real) Ginibre matrix, whose matrix elements have variance $4d\sigma_M^4/N^2=(dN^2)^{-1}$, supported in a disk of radius $1/\sqrt{d}$. If the matrix elements of $K_j$ were normally distributed, then this result would also follow directly from free probability at arbitrary and finite $d$. Since the entries of $K_j$ are \emph{not} normally distributed, we have to resort to the $d\to\infty$ limit and then propose to extend the result to small $d$. This last step is justified \emph{a posteriori} by the remarkable agreement between numerical small-$d$ results and analytical large-$d$ predictions.

In the Appendix, we study the general GinUE-CUE crossover ensemble of matrices of the form $\Phi=a\mathbb{G}+b\mathbb{U}$, $a,b\in\mathbb{R}$, and compute its spectral support and eigenvalue distribution. The results derived there can be readily used to model the quantum map by setting $a=p/\sqrt{d}$ and $b=(1-p)$. 

Since the spectrum of both GinUE and CUE matrices is isotropic in the large-$N$ limit, so is that of the quantum map. In perfect agreement with the numerical results of the previous section, see Fig.~\ref{fig:pKraus_rINOUT}, we find the spectrum to be supported on an annulus whose inner and outer radii are given by 
\begin{equation}\label{eq:pKraus_rpm_analytic}
R_\pm=\frac{1}{\sqrt{d}}\sqrt{(1-p)^2d\pm p^2}.
\end{equation}
The annulus-disk transition in the spectrum occurs at $R_-(p_c)=0$, i.e., $p_c=1/(1+1/\sqrt{d})$. For $p<p_c$, $R_-$ is no longer defined and the spectrum remains a disk (in which case, we conventionalize $R_-\equiv0$). The function $R_+(p)$ is not a monotonic function of $p$, its minimum, $(d+1)^{-1/2}$, being at $p_m=1/(1+1/d)\neq p_c$. The radial eigenvalue distribution, $\varrho(r)=2\pi r\varrho(r,\theta)$, $r\in[R_-,R_+]$, is given by,
\begin{equation}\label{eq:pKraus_rhoR_RMT}
\varrho(r)=2r\frac{d}{p^2}\left(1-\frac{(1-p)^2d}{\sqrt{p^4+4(1-p)^2d^2 r^2}}\right).
\end{equation}
The near-perfect agreement of the analytical radial distribution with the exact-diagonalization results can be seen in Fig.~\ref{fig:pKraus_global_spectrum}~(a)--(e). The small residual deviations are due to finite-$d$ effects (Fig.~\ref{fig:pKraus_global_spectrum} shows data with $d=3$).

\subsection{Steady-State Properties}

The steady state $\rho_\mathrm{SS}$ is the (in general unique) fixed point of the quantum map $\Phi$, $\Phi(\rho_\mathrm{SS})=\rho_\mathrm{SS}$. 
The model described above supports nontrivial (i.e., non-fully mixed) steady states. 
We find that the steady-state properties are similar to those of a random Lindbladian with \emph{non-Hermitian} jump operators~\cite{sa2020RL}. 
This is an important result as it corroborates that the properties of $\rho_\mathrm{SS}$ of non-trivial generic quantum dynamical processes are solely determined by universality arguments.

In order to characterize the steady state, we consider the following measures:
\begin{enumerate}
	\item Steady-state spectrum (steady-state probability distribution, $P_\mathrm{SS}(\lambda) =\Tr [ \delta(\lambda - \rho_\mathrm{SS} ) ] $).
	\item Rényi entropies. We consider the $n$th moment of the eigenvalue distribution through the $n$th Rényi entropy $S_n=S_n(\rho_{\rm SS})$:
	\begin{equation}
	S_n(\rho)=-\frac{1}{n-1}\log\left(\Tr\rho^n\right).
	\end{equation}
	In particular, the first Rényi entropy gives the von Neumann entropy $S_1=-\Tr\left(\rho_\mathrm{SS}\log\rho_\mathrm{SS}\right)$, while the second Rényi entropy is related to the purity of the steady state, $\mathcal{P}_\mathrm{SS}=\Tr\rho_\mathrm{SS}^2=e^{-S_2}$.
	\item Entanglement spectrum. We define an effective Hamiltonian $\mathcal{H}_\mathrm{SS}=-\log\rho_\mathrm{SS}$ and study its spectrum, instead of the spectrum of the steady state itself. Of particular interest are its spectral statistics (e.g., level spacing ratios), that can distinguish an ergodic steady state from a regular one.
\end{enumerate}

In the following, we first analyze the limiting cases of very large ($p=1$) and very small ($p\to0^+$) dissipation separately, exactly determining their steady-state spectral distributions. We then examine the crossover regime interpolating between these two limits, paying special attention to the purity and the correlations in the entanglement spectrum.

\begin{figure}[tp]
    \centering
    \includegraphics[width=\columnwidth]{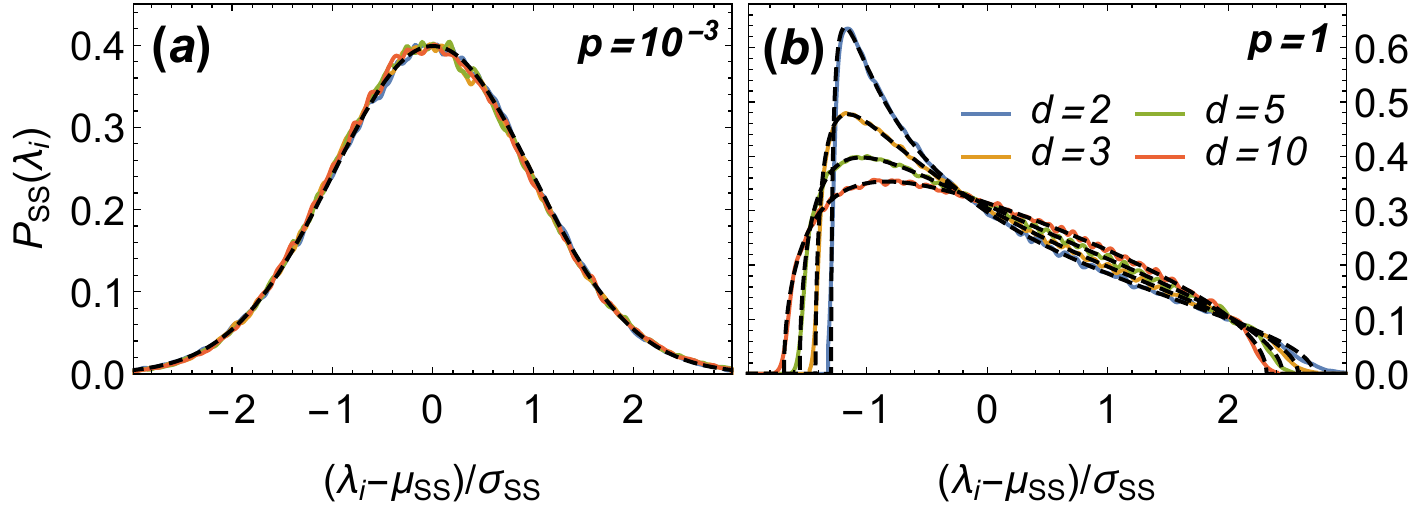}
    \caption{Steady-state eigenvalue distribution (eigenvalues centered at their mean, $\mu_\mathrm{SS}$, and rescaled by their standard-deviation, $\sigma_\mathrm{SS}$, at (a) small and (b) large dissipation, for $N=40$ and $d=2,3,5,10$. Coloured full lines correspond to a smoothed histogram of eigenvalues obtained numerically by exact diagonalization of 5000 steady-state density matrices. The (rescaled and recentered) distributions conform to a normal distribution $\mathcal{N}(0,1)$ at small dissipation and to the Marchenko-Pastur law~(\ref{eq:SS_MP}) at large dissipation (black dashed lines).}
    \label{fig:pKraus_SS_dist}
\end{figure}

\subsubsection{Large dissipation}

At large dissipation, $p\to1$, the steady-state distribution conforms to a Marchenko-Pastur distribution~\cite{marchenko1967}, see Fig.~\ref{fig:pKraus_SS_dist}~(b), in agreement with general results of the entanglement spectrum of random bipartite systems~\cite{zyczkowski2001,sommers2004,znidaric2006,zyczkowski2011,yang2015} (where one takes the partial trace over all environment degrees of freedom, after an infinite-time evolution under joint system-environment unitary dynamics). More precisely, we have to consider a fixed-trace Wishart ensemble to account for probability conservation, $\Tr\rho_\mathrm{SS}=1$. Fixing $\Tr\rho_\mathrm{SS}=1$ leads to a rescaled Marchecko-Pastur probability distribution~\cite{forrester2010,nadal2011}, given by
\begin{equation}\label{eq:SS_MP}
    P_\mathrm{SS}(\lambda)\big\rvert_{p=1}=P_\mathrm{MP}(\lambda;\lambda_\pm)=\frac{1}{2\pi\kappa}\frac{1}{\lambda}\sqrt{(\lambda_+-\lambda)(\lambda-\lambda_-)},
\end{equation}
where $\lambda_\pm=\kappa\left(\sqrt{d}\pm1\right)^2$. The only free parameter, $\kappa$, is fixed by the trace normalization ($\mu_\mathrm{SS}=1/N$), yielding $\kappa=1/(Nd)$, whence $\lambda_\pm=(1/N)(1\pm1/\sqrt{d})^2$. Higher (non-central) moments of the distribution are readily found to be~\cite{mingo2017} 
\begin{equation}\label{eq:moments_Wishart}
    \Tilde{\mu}_n=\frac{\kappa^n}{n}\sum_{\ell=1}^n \binom{n}{\ell-1}\binom{n}{\ell}d^\ell.
\end{equation}
Of particular importance for what follows is the variance $\sigma^2_\mathrm{MP}\equiv\Tilde{\mu}_2-1/N^2=1/(N^2d)$.

\subsubsection{Small dissipation}

At small dissipation, $p\to0^+$, the steady-state eigenvalues are \emph{not} correlated, having instead independent Gaussian distributions, see Fig.~\ref{fig:pKraus_SS_dist}~(a). To show this, we study the steady state perturbatively. At exactly $p=0$, there is an $N$-fold degeneracy of the unit eigenvalue, which gets lifted by any amount of non-unitarity. Nonetheless, we expect the sector of degenerate eigenstates to completely determine the steady-state properties as long as the shift in the eigenvalues due to dissipation is smaller than the typical eigenvalue spacing at unitarity, i.e., for $N p\lesssim 2\pi$. We, therefore, expect a perturbative crossover regime (on the scale $1/N$) towards the Marchenko-Pastur regime. The crossover regime is thus highly suppressed in the thermodynamic limit $N\to\infty$. 

Let $U$ be diagonal in the basis $\{\ket{\alpha}\}$, $\alpha=1,\dots,N$, such that $U\ket{\alpha}=\exp{i\theta_\alpha}\ket{\alpha}$. We evaluate the steady-state-defining equality $\rho_\mathrm{SS}=\Phi(\rho_\mathrm{SS})$ in this basis:
\begin{equation}
\begin{split}
    \rho_{\alpha\beta}=
    &(1-p)\exp{i(\theta_\alpha-\theta_\beta)}\rho_{\alpha\beta}\\
    &+p\sum_{j=1}^d\sum_{\gamma,\delta=1}^N\bra{\alpha}M_j\ket{\beta}\rho_{\gamma\delta}\bra{\delta}M_j^\dagger\ket{\beta}.
\end{split}
\end{equation}
From the preceding discussion, at very small $p$ we can restrict ourselves to the degenerate subspace (with zero phase), i.e., to diagonal elements of $\rho_\mathrm{SS}$. The constant-in-$p$ terms cancel and we obtain
\begin{equation}\label{eq:stochastic_eq}
    \rho_{\alpha\alpha}=
    \sum_{\gamma=1}^N\left(\sum_{j=1}^d\abs{(M_j)_{\alpha\gamma}}^2\right)\rho_{\gamma\gamma}
    \equiv \sum_{\gamma=1}^N T_{\alpha\gamma}\rho_{\gamma\gamma}.
\end{equation}
This equation has the immediate interpretation of a classical probability equation: the diagonal elements of $\rho_\mathrm{SS}$ form the invariant probability measure of the random stochastic matrix $T$~\cite{zyczkowski2003,horvat2009,chafai2010}. That $T$ is a stochastic matrix, i.e.
\begin{subequations}\label{eq:stochastic_eq_conditions}
\begin{align}
    \mathbb{R}\ni T_{\alpha\gamma}\geq0,\\
    \sum_{\alpha=1}^N T_{\alpha\gamma}=1,
\end{align}
\end{subequations}
follows immediately from its definition in Eq.~(\ref{eq:stochastic_eq}) and from the orthonormality of rows and columns of $V$ (recall from Sec.~\ref{sec:model} that the $M_j$ are truncations of the unitary $V$). The distribution of the entries of $T$ can also be immediately inferred. Given that $(M_j)_{\alpha\gamma}=(V_{j1})_{\alpha\gamma}$ are entries of a $(Nd\times Nd)$ Haar-random unitary, which are known to be complex-normal distributed~\cite{pereyra1983}, the entries of $T$ are the sum of the squares of $2d$ real normal-distributed random variables with zero mean and variance $1/(2Nd)$. Therefore, $(2Nd)T_{\alpha\gamma}$ follows a $\chi^2$-distribution with $2d$ degrees of freedom. Note that the matrix elements $T_{\alpha\gamma}$ thus have mean $1/N$, as required from Eq.~(\ref{eq:stochastic_eq_conditions}).

By the Perron-Frobenius theorem~\cite{bengtsson2017}, the maximal eigenvalue of $T$ is $1$ and the corresponding eigenvector (the invariant probability measure, or the diagonal entries of $\rho_\mathrm{SS}$ in our case) has real non-negative entries.\footnote{
The conditions for the Perron-Frobenius theorem to hold, namely that $M$ is irreducible and aperiodic are met almost surely because the entries of $M$ are, with probability one, nonzero (since they are the squares of the entries of a Haar-random unitary).
}

We now show that the steady-state probabilities of a random stochastic matrix are normally-distributed, following Ref.~\cite{horvat2009}. 
We assume that, in the $N\to\infty$ limit, $\rho_{\alpha\alpha}$ and $T_{\alpha\gamma}$ become independent random variables. Then, for fixed $\alpha$ and $\gamma$, $T_{\alpha\gamma}\rho_{\gamma\gamma}$ (no sum over $\gamma$) has a product distribution. We denote the distributions of $\rho_{\alpha\alpha}$, $T_{\alpha\gamma}$, and $T_{\alpha\gamma}\rho_{\gamma\gamma}$ by $P_\rho$, $P_T$, and $P_{T\rho}$, respectively. The mean and variance of these distributions are $\mu_\rho$, $\sigma_\rho^2$, etc. The product-distribution moments satisfy $\mu_{T\rho}=\mu_T\mu_\rho$ and
\begin{equation}\label{eq:sigma_product}
    \sigma_{T\rho}^2=\left(\sigma_T^2+\mu_T^2\right)\left(\sigma_\rho^2+\mu_\rho^2\right)-\mu_T^2\mu_\rho^2.
\end{equation}
Now, $\rho_{\alpha\alpha}$ is the sum of $N$ such independently distributed matrices [recall Eq.~(\ref{eq:stochastic_eq})] and, when $N\to\infty$, by the central limit theorem, it is normally-distributed, $\rho_{\alpha\alpha}\sim\mathcal{N}(\mu_\rho,\sigma_\rho)$, with $\mu_{\rho}=N\mu_{T\rho}$ and $\sigma^2_\rho=N\sigma^2_{T\rho}$. This procedure turned the stochastic steady-state equation into a self-consistent condition fixing $\sigma_\rho^2$: the Gaussian distribution of $\rho$ is completely determined by the first two moments of $P_{T\rho}$, which in turn depend only on the two lowest moments of $P_T$ and on $P_\rho$ itself. Substituting $\sigma^2_\rho=N\sigma^2_{T\rho}$ into Eq.~(\ref{eq:sigma_product}), we find
\begin{equation}\label{eq:sigma_rho}
    \sigma^2_\mathrm{P}\equiv\sigma^2_\rho
    =\frac{\sigma^2_T}{N}\frac{1}{1-N\sigma^2_T-\frac{1}{N}}
    =\frac{1}{N^3d}\frac{1}{1-\frac{1}{N}\left(1+\frac{1}{d}\right)},
\end{equation}
where we used $\sigma^2_T=1/(N^2d)$ for the $\chi^2$-distributed random variable $T_{\alpha\gamma}$ and the subscript P distinguishes this perturbative variance from the Marchenko-Pastur variance, $\sigma^2_\mathrm{MP}$, obtained above.

The classical-probability-equation structure of the quantum dynamical equation resulting from perturbation theory in the degenerate subspace at small dissipation was already identified in Ref.~\cite{sa2020RL} (at the level of the continuous-time classical Markov generator) and used to study the spectral gap of a random Liouvillian; however, its steady-state properties were not investigated. From these results (see also the related Refs.~\cite{timm2009,horvat2009}), we see that, at small deviations from unitarity, both spectral and steady-state properties are found to depend solely on the first two moments of a random matrix of small size (i.e., of order $N$ instead of $N^2$).

\begin{figure}[tp]
    \centering
    \includegraphics[width=\columnwidth]{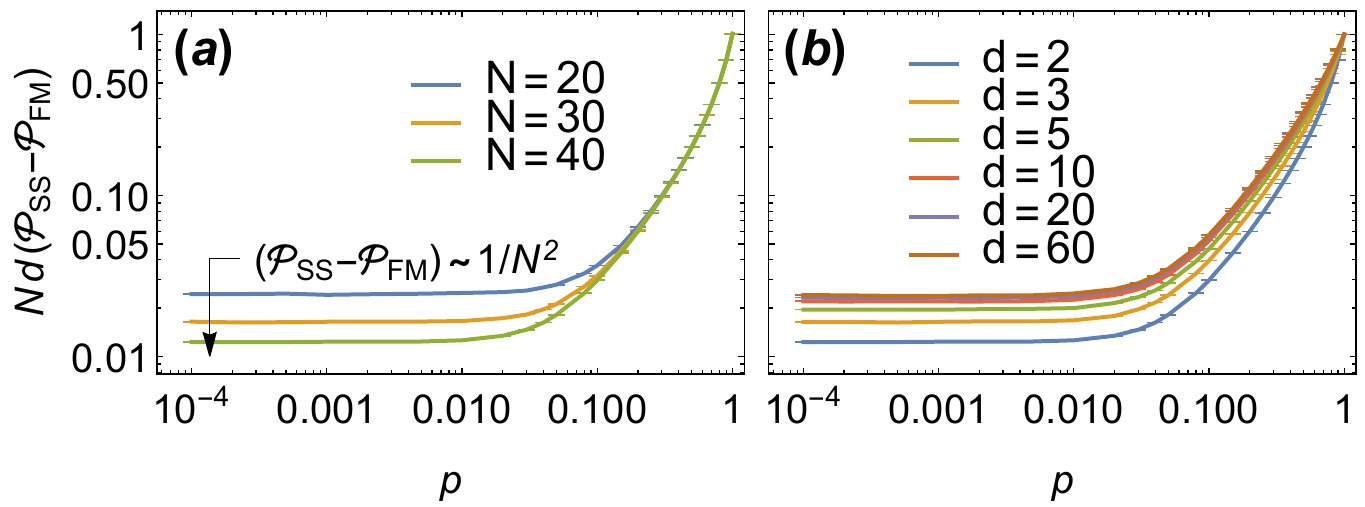}
    \caption{Difference between the steady-state purity and the purity of the fully-mixed state, as a function of $p$. Numerical results obtained by exact diagonalization of 5000 steady-state density matrices. (a): fixed $d=2$ and varying $N$. The purity difference scales as $1/N$ for large-enough $p$ and with $1/N^2$ in the small-$p$ perturbative regime. (b): fixed $N=40$ and varying $d$. For large $d$, the purity scales as $1/d$ across all dissipation regimes (the curves collapse to a single universal curve for all $p$).}
    \label{fig:pKraus_SS_purity}
\end{figure}

\subsubsection{Purity}

Figure~\ref{fig:pKraus_SS_purity} shows the difference $\mathcal{P}_\mathrm{SS}-\mathcal{P}_\mathrm{FM}$ between the steady-state purity, $\mathcal{P}_\mathrm{SS}$, and that of the fully mixed state, $\mathcal{P}_\mathrm{FM}=1/N$, as a function of $p$ (also note that $\mathcal{P}_\mathrm{SS}-\mathcal{P}_\mathrm{FM}=N\sigma_\mathrm{SS}^2$). By rescaling the purity difference by $1/N$, curves of different $N$ collapse to a universal curve in the large-$p$ limit. At exactly $p=1$, Eq.~(\ref{eq:moments_Wishart}), gives $\av{\mathcal{P}_\mathrm{SS}-\mathcal{P}_\mathrm{FM}}=N\sigma^2_\mathrm{MP}=\mathcal{P}_\mathrm{FM}/d$. This scaling holds for a finite range of $p$, but as $p$ is decreased, the individual curves depart from the universal curve and enter the perturbative crossover regime, characterized by a purity difference proportional to $1/N^2$, as follows from Eq.~(\ref{eq:sigma_rho}) in the large-$N$ limit. Thus, in the small-$p$ regime, the steady state can be considered fully mixed, since $\av{\mathcal{P}_\mathrm{SS}-\mathcal{P}_\mathrm{FM}}\ll \mathcal{P}_\mathrm{FM}$. 

In the thermodynamic limit, the universal Marchenko-Pastur curve covers the entire range of $p$, the fully-mixed state, $\mathcal{P}_\mathrm{SS}-\mathcal{P}_\mathrm{FM}=0$, being achieved only at exactly $p=0$. Also in the thermodynamic limit, the purity scales as $1/d$ for all dissipation strengths, at least for large enough $d$ (see Fig.~\ref{fig:pKraus_SS_purity}~(b)).

\begin{figure}[tp]
    \centering
    \includegraphics[width=\columnwidth]{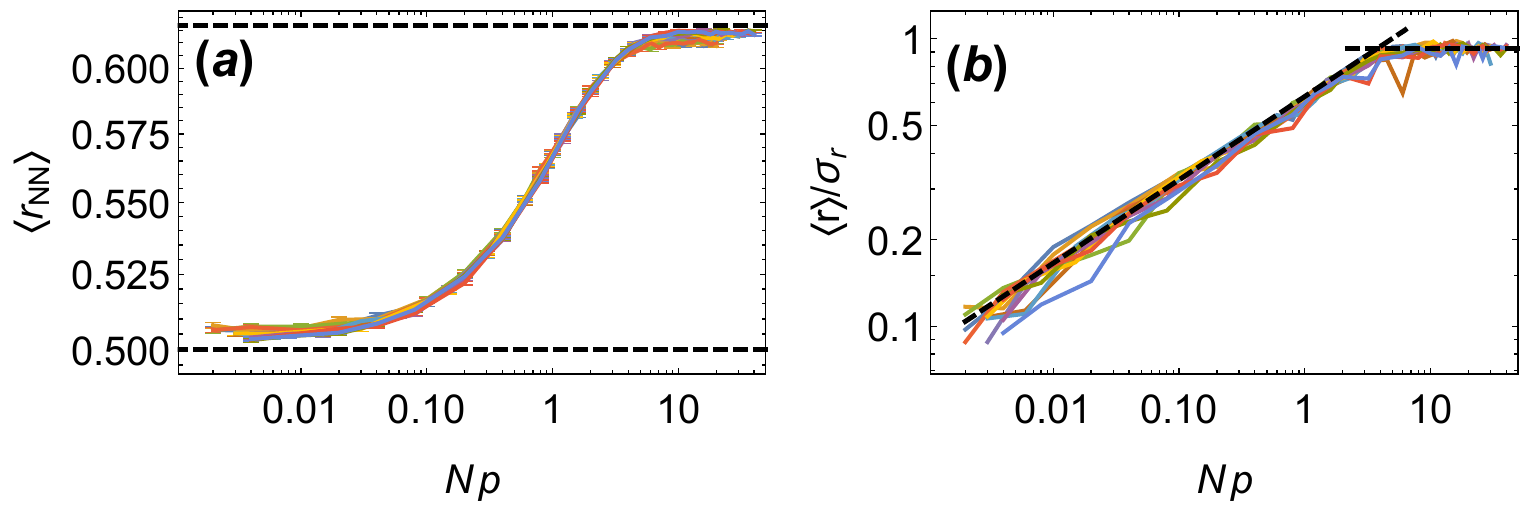}
    \caption{Level spacing statistics of the entanglement Hamiltonian as a function of $Np$. (a): average of the NN/NNN ratio. The dashed lines correspond to the theoretical values for Poisson statistics ($\av{\abs{r_\mathrm{NN/NNN}}}=1/2$, exact) and RMT statistics ($\av{\abs{r_\mathrm{NN/NNN}}}\approx 0.617$, approximate). (b): ratio of the first two moments of the distribution of the consecutive level spacing ratio, $\av{r}/\sigma_r$. The horizontal dashed line corresponds to the theoretical RMT value for the unitary symmetry class ($\av{r}/\sigma_r\approx0.928$), while the other dashed line gives the power-law decrease towards the Poisson limit ($\av{r}/\sigma_r=0$). The curves of different colors correspond to all the combinations of $N=30,40,50$ and $d=2,3,5,10$, collapsed to a single universal curve, clearly showing the crossover regime to scale as $p\sim1/N$.}
    \label{fig:pKraus_SS_cross}
\end{figure}

\subsubsection{Perturbative crossover}

The perturbative crossover is best seen in the spectral statistics of the entanglement Hamiltonian $\mathcal{H}_\mathrm{SS}$, which are captured by level-spacing statistics~\cite{guhr1998}. Furthermore, to automatically unfold the spectrum, we consider ratios of spacings. We denote the eigenvalues of $\mathcal{H}_\mathrm{SS}$ by $\varepsilon_i$. The nearest-neighbour to next-to-nearest-neighbour (NN/NNN) spacing ratio~\cite{sa2019CSR,srivastava2018} is defined by $r_\mathrm{NN/NNN}=(\varepsilon^\mathrm{NN}_i-\varepsilon_i)/(\varepsilon^\mathrm{NNN}_i-\varepsilon_i)$, where $\varepsilon^\mathrm{NN(NNN)}_i$ denotes the level nearest (next-to-nearest) to $\varepsilon_i$. In Fig.~\ref{fig:pKraus_SS_cross}~(a), we show the average value of $\abs{r_\mathrm{NN/NNN}}$ as a function of dissipation strength, which clearly distinguishes the Poisson statistics of uncorrelated levels ($\av{\abs{r_\mathrm{NN/NNN}}}=1/2$) from the random matrix statistics in the unitary class (a Wigner-like surmise gives an approximate value of $\av{\abs{r_\mathrm{NN/NNN}}}\approx0.617$~\cite{sa2019CSR}). As the size of the system increases, there is no sharp transition between both limits. Alternatively, we can consider the ratios of consecutive spacings~\cite{oganesyan2007,atas2013,atas2013long}, $r_i=(\varepsilon_{i+1}-\varepsilon_i)/(\varepsilon_i-\varepsilon_{i-1})$. For uncorrelated levels, the distribution of $r$, which can be computed exactly, $P_r(r)=1/(1+r)^2$, has all moments but the first undefined (diverging), whence the ratio of the first two moments $\av{r}/\sigma_r$ is zero. On the other hand, for random matrix statistics, the latter is given by a finite value of order one (a Wigner-like surmise calculation gives the result $\av{r}/\sigma_r\approx0.928$~\cite{atas2013}). This is illustrated in Fig.~\ref{fig:pKraus_SS_cross}~(b): outside the perturbative regime, $Np\gtrsim2\pi$, $\av{r}/\sigma_r$ has the RMT value, and there is a power-law crossover to $\av{r}/\sigma_r=0$, which is strictly attained only for $Np=0$. The exponent of the power-law decay, $\av{r}/\sigma_r\sim (Np)^\alpha$, is numerically found to be $\alpha\approx0.3$.

From the discussion above, we conclude that no signature of the spectral transition, at finite $p$, is imprinted on the steady state. On the contrary, the steady-state properties are highly universal, and strongly resembling those of random Lindbladians once the different parametrizations of non-unitarity are taken into account.

\section{1D Random Kraus Circuits}
\label{sec:1d}

\begin{figure}[tp]
	\centering
	\includegraphics[width=\columnwidth]{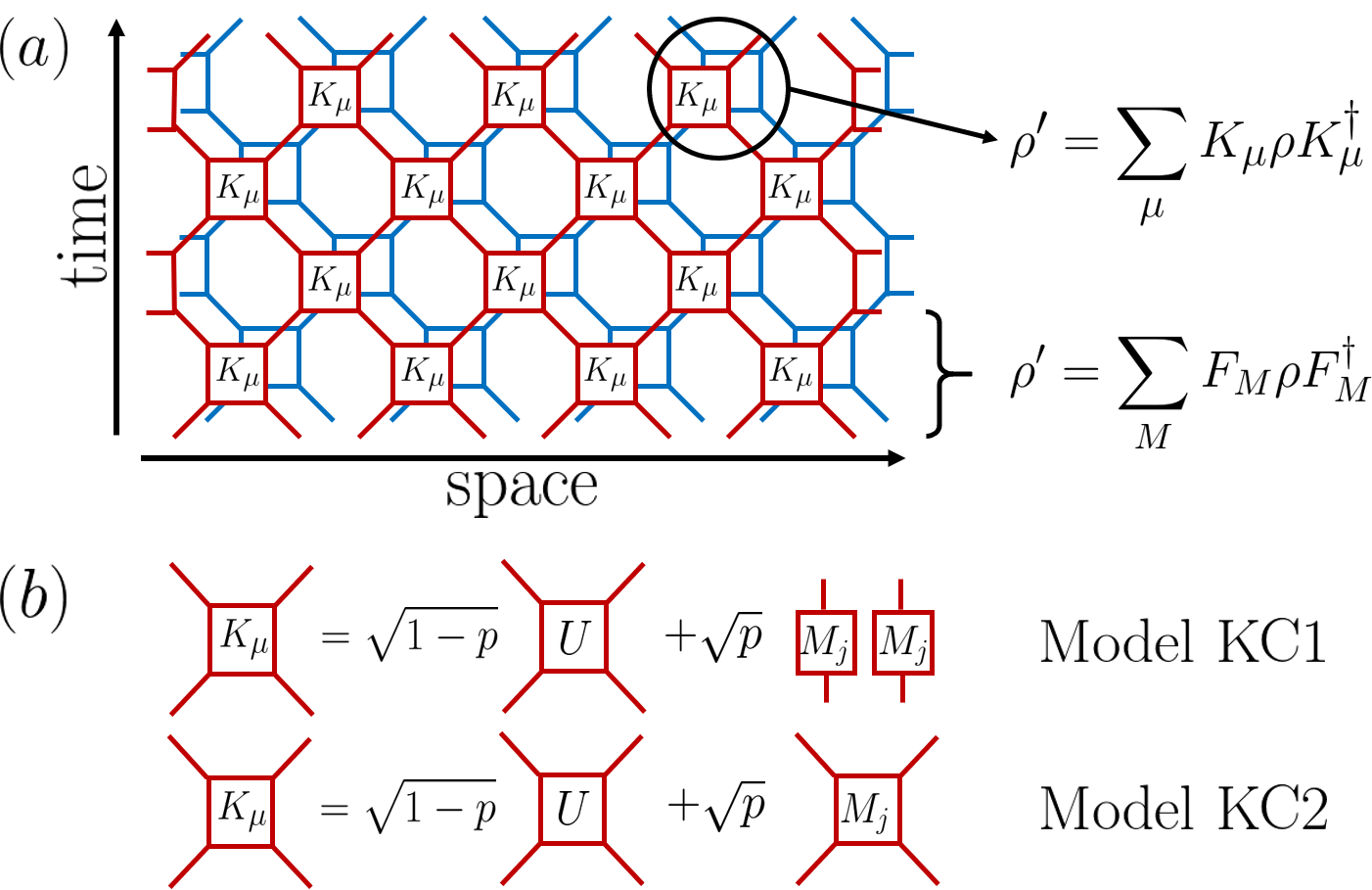}
	\caption{1D local Kraus circuit. (a): ``space-time'' structure of the quantum circuit. The red and blue layers represent the two copies of the system (i.e., one tensor-product factor each in the matrix representation). A local Hilbert space $\mathcal{H}_j$, of dimension $q$, lives on each wire connecting two Kraus operators $K_\mu$, represented by a four-legged square. A full row of the circuit is described by Kraus operators $F_M$, while one time-step of this Floquet Kraus circuit is given by two rows. Periodic boundary conditions are imposed. (b): each ``brick'' represents a two-body Kraus map with Kraus operators of the form of Eq.~(\ref{eq:Kraus_def}). We consider two models (both with two-body unitary dynamics) differing in whether the dissipative contribution is a genuine two-body Kraus operator (model KC2) or factors into one-body operators (KC1).}
	\label{fig:kraus_circuit}
\end{figure}

Next, we consider a one-dimensional system of $L$ qudits of dimension $q$, with local Hilbert space $\mathcal{H}_j=\mathbb{C}^{q}$. For convenience $L$, is taken to be even. 
The time-evolution superoperator corresponds to two rows of the Kraus \emph{circuit} schematically depicted in Fig.~\ref{fig:kraus_circuit}~(a). Note that the same local two-site Kraus map is applied everywhere along the space and time directions.
The composition of two subsequent rows (even and odd) yields one time-step of the Floquet Kraus dynamics. 
One row of the quantum circuit (half a time-step), $\Phi$, can be written in terms of global Kraus operators $F_M$, $\Phi(\rho)=\sum_M F_M\rho F_M^\dagger$, where $M=(\mu_1,\mu_2,\dots,\mu_{L/2})$ is a multi-index, which, in our model, factorize into local two-body Kraus operators $K_{\mu_j}$, of the form of Eq.~(\ref{eq:Kraus_def}), $F_M=K_{\mu_1}\otimes\cdots\otimes K_{\mu_{L/2}}$. The full one-time-step quantum map (two rows of the circuit) is correspondingly given by $\mathbb{T}\Phi\mathbb{T}^\dagger \Phi$, where we have introduced the one-site translation operator $\mathbb{T}$, defined by its action on the computational-basis states:
\begin{equation}
\mathbb{T}\ket{s_1,s_2,\dots,s_L}=\ket{s_L,s_1,\dots,s_{L-1}}, 
\end{equation}
with the quantum number $s_j$ being the spin-$q$ of each site.
We consider two models, both with two-site unitary dynamics, but differing in the number of sites on which the dissipative Kraus operators act: in model KC1, the dissipative two-body Kraus operator $M_j$ factorizes into two one-body Kraus operators $M_{j_1}\otimes M_{j_2}$, while in model KC2 $M_j$ is a genuine two-body operator, see Fig.~\ref{fig:kraus_circuit}~(b).

\begin{figure*}[tp]
    \centering
    \includegraphics[width=0.93\textwidth]{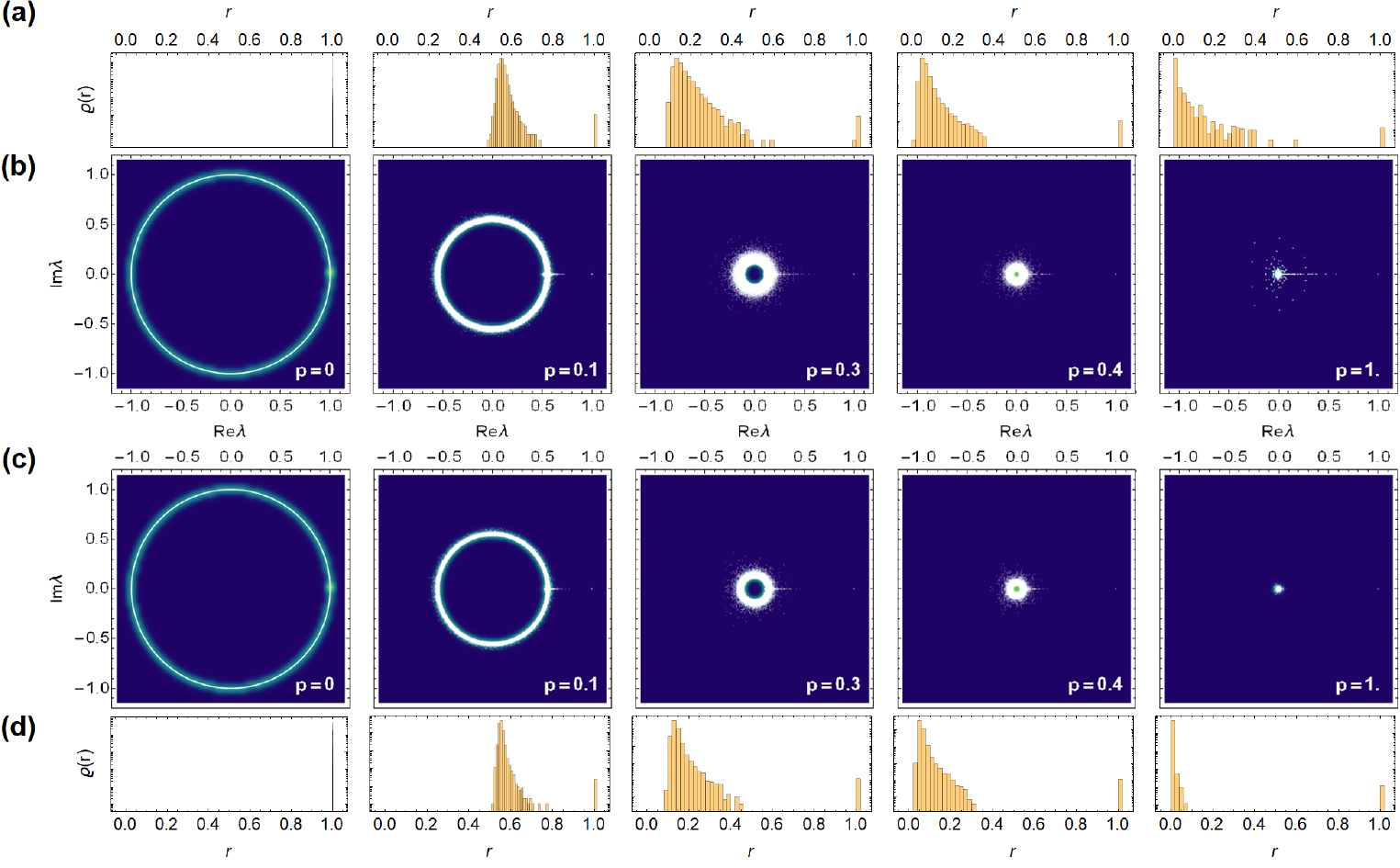}
    \caption{Spectral density of the one-dimensional Kraus circuit ($L=6$ and $q=2$), for different values of $p\in[0,1]$. (a) and (b): model KC1 with $d=3$; (c) and (d): model KC2 with $d=15$. (The radial histograms in (a) and (d) have a logarithmic scale.) The qualitative features (including annulus-disk spectral transition) of the spectra of Fig.~\ref{fig:pKraus_global_spectrum} are also present here. Note that while for the global map the spectrum has sharp and well-defined boundaries (inner and outer radii), this is no longer the case here, with several isolated eigenvalues lying outside a more diffusive boundary (a scatter plot of the individual eigenvalues is superimposed on the spectral density). Note also that the circuit is more contractive here because we introduce interactions by applying \emph{two} layers of quantum maps.}
    \label{fig:pKC_global_spectrum}
\end{figure*}

\begin{figure}[tp]
    \centering
    \includegraphics[width=0.96\columnwidth]{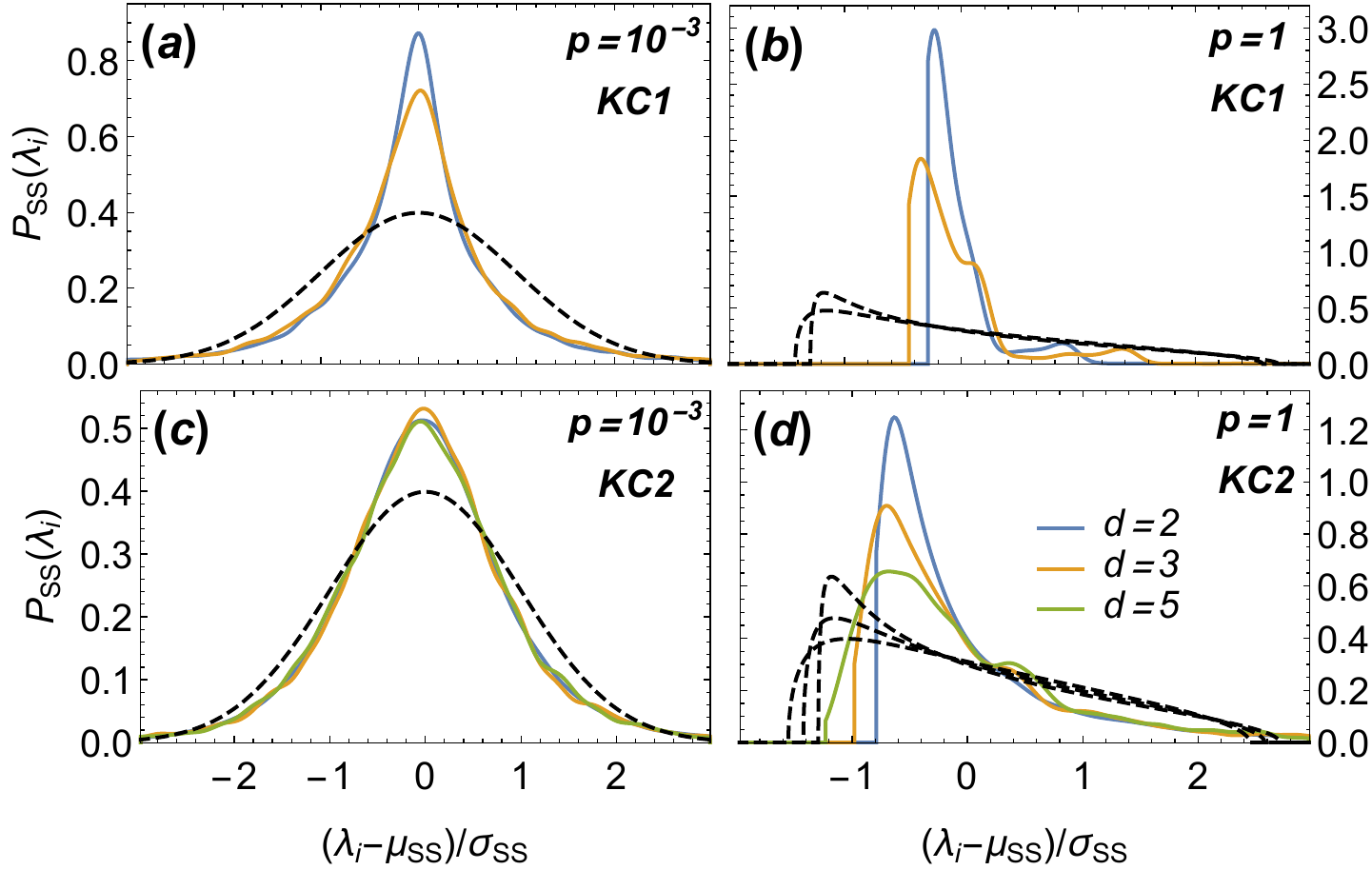}
    \caption{Steady-state eigenvalue distribution (eigenvalues centered at their mean, $\mu_\mathrm{SS}$, and rescaled by their standard-deviation, $\sigma_\mathrm{SS}$), for $L=6$ and $d=2,3,5$. We show results for model KC1 [KC2] at (a) [(c)] small and (b) [(d)] large dissipation. The qualitative features are the same as in Fig.~\ref{fig:pKraus_SS_dist} are still present, although an exact description in terms of Gaussian (a,c) or Marchenko-Pastur (b,d) distributions is no longer possible (see dashed lines).}
    \label{fig:pKC_SS_dist}
\end{figure}

For the local Kraus circuit, we study the same quantities analyzed for the fully-connected quantum map. Quite remarkably, some of the properties of the local version are qualitatively similar to those of the global map. Furthermore, both models KC1 and KC2 also display essentially the same features. This indicates that the statistical properties of chaotic quantum maps are quite generic.

Figure~\ref{fig:pKC_global_spectrum} shows the spectrum of the quantum maps $\mathbb{T}\Phi\mathbb{T}^\dagger \Phi$ for $q=2$ and $L=6$, increasing $p$, and both KC1 ($d=3$) and KC2 ($d=15$). As before, the spectrum evolves from being supported on the unit circle, at $p=0$ to an annulus for $0<p<p_c$, and then undergoing a transition at $p=p_c$ to a disk spectral support. While this qualitative behavior is exactly that of the fully-connected model, the effective RMT model discussed in Sec.~\ref{subsec:RMT_model} does not quantitatively describe the spectral density, because the map is built from tensoring several maps of \emph{small} dimension. In particular, while the boundaries of the support were very sharp before, there are now several isolated eigenvalues lying outside a diffuse boundary (this can be seen in Fig.~\ref{fig:pKC_global_spectrum}, where we have superimposed a scatter plot of the individual eigenvalues on the spectral density). Besides, the spectrum is not even approximately flat inside its support.

Another qualitative difference from the structureless random Kraus maps is the appearance of a (zero-measure) set of eigenvalues along the real positive axes. 
This feature can be understood as follows. General hermiticity-preserving operators (such as $\Phi$) can be brought to a real representation by a trivial similarity transformation. Therefore, the dissipative part of the map should be modelled by a real Ginibre matrix (drawn from the GinOE), instead of by a complex Ginibre matrix (drawn form the GinUE).\footnote{
However, because the bulk distribution and the correlation statistics are the same for the GinUE and the GinOE and the former is considerably easier to work with, we conveniently modelled $\mathbb{G}$ by a GinUE matrix without much loss of accuracy.}
Now, the matrices from the GinOE have nonzero spectral weight on the real axis, but it is suppressed in the large-$N$ limit~\cite{forrester2007}. Therefore, for the 0D map, where the convergence to the large-$N$ limit is faster, the real spectral weight is strongly surpassed. In contrast, for the 1D circuit, spectral weight along the real line is still visible within the system sizes available to us. This is so because $\Phi$ is constructed by tensoring together many maps of small dimension---each with a finite real spectral weight---which attain the large-$N$ limit in a slower fashion. Finally, the spectral gap is determined by the largest of these real eigenvalues and remains finite in the thermodynamic limit for $p<0$.

We now turn to the steady state of the random Kraus circuit. In Fig.~\ref{fig:pKC_SS_dist} we show the spectrum of the steady state for small ($p=10^{-3}$) and large $(p=1)$ dissipation. As before, at small $p$, the steady-state spectrum has a symmetric peak around $1/N$, with a variance that is again found to scale as $1/N^3$. For large $p$, the distribution is no longer symmetric, there being a longer tail on the right, and the variance scales with $1/N^2$. 
Even though the main qualitative features of the 0D fully-connected model are still present, it is important to remark that the steady-state spectrum is \emph{not} accurately described by a normal distribution (Marchenko-Pastur law), at small (large) dissipation, see Fig.~\ref{fig:pKC_SS_dist}. These deviations, most notably the bumps in the distribution at large $p$, should be attributed to the small dimension of the local Hilbert space we could access, hindering an exact description in terms of large-$N$ RMT.

\hfill

\section{Conclusions}
\label{sec:conclusion}

We have analyzed the spectral and steady-state properties of two kinds of stochastic Floquet non-unitary dynamics: a 0D global (or spatially structureless) Kraus map and a 1D circuit composed of local Kraus operators. 
By changing a single parameter $p\in[0,1]$, loosely regarded as the dissipation strength, the family of Kraus maps we consider interpolates between a random unitary operator, at $p=0$, and a generic quantum stochastic map with $d$ channels, at $p=1$.    

Although qualitatively similar, the 0D and 1D cases show some important differences.
For 0D, all the spectral weight (except for the eigenvalue one corresponding to the steady state) is supported either on an annulus, for $p<p_c$, or on a disk, for $p>p_c$. 
We determined the exact eigenvalue density using an ansatz that perfectly describes the numerical data.
The spectral gap, which determines the asymptotic decay rate to the steady state, coincides with the outer radius of the spectral support and shows a curious non-monotonic behavior as a function of $p$. 
For 1D, the spectrum depicts the same overall features at low (annulus) and large (disk) values of $p$.  
However, the support of the spectral weight is not sharply defined, rendering the annulus-ring transition difficult to determine numerically with the available system sizes. The spectral gap is determined by a set of eigenvalues of vanishing relative spectral weight laying along the real positive axis. 

Regarding the properties of the steady state, the 0D and 1D cases are again qualitatively very similar. For the 0D case, at low dissipation, the steady-state eigenvalue distribution is Gaussian, whereas it conforms to a Marchenko-Pastur law at large dissipation. These features are in correspondence with a Poissonian (small $p$) or GUE (large $p$) level-spacing distributions. 
The crossover between these two regimes is captured by a scaling function of $N p$. Therefore, in the thermodynamic limit, an infinitesimal amount of dissipation renders the spectrum of the steady-state Marchenko-Pastur distributed with GUE level-spacing statistics. 
The fact that the same behavior has previously been found for random Lindblad operators~\cite{sa2020RL} indicates that such steady-state features are to be generically expected for stochastic open quantum systems. 

Our analysis shows that, although some global features are similar, the unstructured or local nature of the evolution operators imprints characteristic signatures in the spectrum.
Whether these are present in more realistic physical models is an important question for further studies.  It would also be interesting to search for signatures of the annulus-ring transition on the dynamics. 

Finally, our results point to the existence of rather universal steady-state properties of stochastic Markovian dissipative models. Here, again, further work is needed to determine whether these are also present in realistic models of open quantum systems. 

\begin{acknowledgments}
LS acknowledges support by FCT through PhD Scholarship SFRH/BD/147477/2019. PR acknowledges support by FCT through the Investigador FCT contract IF/00347/2014 and Grant No. UID/CTM/04540/2019.
TP acknowledges ERC Advanced grant 694544-OMNES and ARRS research program P1-0402.	
\end{acknowledgments}

\renewcommand{\appendixname}{APPENDIX}
\appendix

\section*{Appendix: G\texorpdfstring{\lowercase{in}}{in}UE-CUE Crossover}

In this Appendix, we compute the spectral support and eigenvalue density for the general GinUE-CUE crossover ensemble
\begin{equation}\label{eq:RMT_model}
\Phi=a\mathbb{G}+b\mathbb{U},
\end{equation}
where $\mathbb{G}$ is a GinUE matrix, $\mathbb{U}$ a CUE matrix, and $a$, $b$ real constants. To this end, we employ quaternionic free probability~\cite{feinberg1997a,feinberg1997b,janik1997a,janik1997b,feinberg2001,janik2001,jarosz2004,jarosz2006,feinberg2006,burda2011,burda2015,nowak2017jstat,nowak2017pre,denisov2018}, which we start by briefly reviewing below.

\subsection{Non-Hermitian free probability review}
\label{sec:RMT_model_analytics}

The quaternionic resolvent (Green's function) of a random matrix $\phi$ is defined by
\begin{equation}
\mathcal{G}(Q)=\av{\frac{1}{N}\mathrm{bTr}(Q-\mathcal{H})^{-1}},
\end{equation}
where $Q$ is a general quaternion parametrized as 
\begin{equation}
Q=\begin{pmatrix}
\alpha & \beta \\
-\conj{\beta} & \conj{\alpha}
\end{pmatrix},
\end{equation}
$\alpha,\beta\in\mathbb{C}$, $\mathcal{H}=\mathrm{diag}(\phi,\phi^\dagger)$, and the block trace $\mathrm{bTr}$ is a partial trace over the Hilbert space variables, returning a $2\times2$ matrix. When inside the block trace, a quaternion Q is to be understood as $Q\otimes \mathbbm{1}$, where $\mathbbm{1}$ is the $N$-dimensional identity matrix. The Green's function is also a quaternion which we parametrize as
\begin{equation}
\mathcal{G}=\begin{pmatrix}
\mathcal{G}_{11} & \mathcal{G}_{12}\\
-\conj{\mathcal{G}}_{12} & \conj{\mathcal{G}}_{11}
\end{pmatrix}.
\end{equation}

Several quantities related to the Green's function prove useful in the following. The functional inverse of the Green's function is the Blue's function $\mathcal{B}(\mathcal{G}(Q))=\mathcal{G}(\mathcal{B}(Q))=Q$, which is related to the $\mathcal{R}$-transform by $\mathcal{R}(Q)=\mathcal{B}-Q^{-1}$. The self-energy $\Sigma(Q)$ is defined as usual by $\mathcal{G}(Q)=(Q-\Sigma(Q))^{-1}$. It then follows that the $\mathcal{R}$-transform evaluated on the Green's function is nothing but the self-energy, $\Sigma(Q)=\mathcal{R}(\mathcal{G}(Q))$.

\hfill
\onecolumngrid

Next, we recall the scaling properties of the Green's and Blue's functions. If $K$ is some quaternion, we have
\begin{equation}\label{eq:scaling_G}
\mathcal{G}_{K\mathcal{H}}(Q)=\mathcal{G}_\mathcal{H}(K^{-1}Q)K^{-1}.
\end{equation}
The Blue's function and self-energy scale inversely:
\begin{equation}\label{eq:scaling_B}
\mathcal{B}_{K\mathcal{H}}(Q)=K\mathcal{G}_\mathcal{H}(QK),
\end{equation}
and identically for the $\mathcal{R}$-transform.

Turning to the sum of two random matrices $A+B$, one can prove~\cite{nica2006,mingo2017} that the $\mathcal{R}$-transform satisfies the additive property $\mathcal{R}_{A+B}(Q)=\mathcal{R}_A(Q)+\mathcal{R}_B(Q)$. From this property one derives the following non-Hermitian Pastur equation:
\begin{equation}
\mathcal{G}_B\left[Q-\mathcal{R}_A\left(\mathcal{G}_{A+B}(Q)\right)\right]=\mathcal{G}_{A+B}(Q).
\end{equation}
At the end of the calculations, we are interested in returning to the complex plane and hence set $\alpha=z\in\mathbb{C}$, $\beta=0$, and obtain
\begin{equation}\label{eq:nonhermitian_Pastur}
\mathcal{G}_B\left[
\begin{pmatrix} 
z & 0 \\ 0 & \conj{z}
\end{pmatrix}
-\mathcal{R}_A\left[
\begin{pmatrix}
\mathcal{G}_{11}(z,\conj{z}) &
\mathcal{G}_{12}(z,\conj{z})\\
-\conj{\mathcal{G}}_{12}(z,\conj{z}) &
\conj{\mathcal{G}}_{11}(z,\conj{z})\\
\end{pmatrix}
\right]
\right]=
\begin{pmatrix}
\mathcal{G}_{11}(z,\conj{z}) &
\mathcal{G}_{12}(z,\conj{z})\\
-\conj{\mathcal{G}}_{12}(z,\conj{z}) &
\conj{\mathcal{G}}_{11}(z,\conj{z})\\
\end{pmatrix}.
\end{equation}

Finally, to obtain the spectral density, we differentiate the upper-left block of the quaternionic Green's function
\begin{equation}
\varrho(z,\conj{z})=\frac{1}{\pi}\partial_{\conj{z}}\mathcal{G}_{11}(z,\conj{z}).
\end{equation}

In our effective RMT model of the quantum map, the matrix $A$ is a GinUE matrix, while $B$ is drawn form the CUE. Accordingly, we need the $\mathcal{R}$-transform of the GinUE and the quaternionic Green's function of the CUE.

\subsection{Quaternionic \texorpdfstring{$\mathcal{R}$}{R}-transform for the GinUE}

The $\mathcal{R}$-transform of the GinUE has been obtained in Refs.~\cite{janik1997a,janik1997b,feinberg1997a} by a variety of different methods. For completeness, we briefly lay out the computation starting from the (Hermitian) Gaussian Unitary Ensemble (GUE). A matrix $A$ from the GinUE can be parametrized in terms of two Hermitian matrices $H$, $H'$ from independent Gaussian ensembles, $A=(H+iH')/\sqrt{2}$. The defining feature of the Gaussian ensembles is the additivity of its Green's functions (i.e., the sum of Gaussian matrices is still Gaussian, the result for random matrices analogous to the central limit theorem for classical probability), which implies $\mathcal{G}=\Sigma$ or, equivalently, $\mathcal{R}_H(Q)=Q$. This fact, together with the additivity of the $\mathcal{R}$-transform and the scaling relation~(\ref{eq:scaling_B}), yields
\begin{equation}
\mathcal{R}_A=\mathcal{R}_{H/\sqrt{2}}+\mathcal{R}_{iH'/\sqrt{2}}=\frac{1}{2}(Q+\mathcal{I}Q\mathcal{I}),
\end{equation} 
where $\mathcal{I}=\mathrm{diag}(i,-i)$. We then immediately obtain the $\mathcal{R}$-transform for the GinUE,
\begin{equation}\label{eq:G_GinuE}
\mathcal{R}_\mathrm{GinUE}(Q)=
\begin{pmatrix}
0 & \beta \\
-\conj{\beta} & 0
\end{pmatrix}.
\end{equation}

\subsection{Quaternionic Resolvent for the CUE}

We now consider the quaternionic resolvent for the CUE. Related results were given in Refs.~\cite{jarosz2011,nowak2017pre}, but the end-result of this computation, Eq.~(\ref{eq:G_CUE}), is not explicitly given in the literature, to the best of our knowledge. Because $\mathbb{U}$ is normal (i.e., $\comm{\mathbb{U}}{\mathbb{U}^\dagger}=0$), its left- and right-eigenvectors coincide. We can then write
\begin{equation}
\mathbb{U}=\sum_{n}\ket{n}e^{i\theta_n}\bra{n}
\qquad \text{and}\qquad
\mathbb{U}^\dagger=\sum_{n}\ket{n}e^{-i\theta_n}\bra{n},
\end{equation}
in some eigenbasis $\{\ket{n}\}$. The argument of the Green's function $\mathcal{G}(Q)$ can then be easily inverted,
\begin{equation}
(Q-\mathcal{H})^{-1}=\sum_n\ket{n}\frac{1}{(\alpha-e^{i\theta_n})(\conj{\alpha}-e^{-i\theta_n})+\conj{\beta}\beta}
\begin{pmatrix}
\conj{\alpha}-e^{-i\theta_n} & -\beta \\
\conj{\beta} & \alpha-e^{i\theta_n}
\end{pmatrix}
\bra{n}.
\end{equation}

Taking the large-$N$ limit, we can replace $\av{(1/N)\sum_nF(\theta_n)}$ by $\int \d\theta \varrho(\theta)\av{F(\theta)}$, where $\varrho(\theta)=1/(2\pi)$ is the flat density of the CUE on the unit circle. Performing the block trace, the quaternionic resolvent reads
\begin{equation}
\mathcal{G}(Q)=\int\frac{\d\theta}{2\pi}\frac{1}{(\alpha-e^{i\theta})(\conj{\alpha}-e^{-i\theta})+\conj{\beta}\beta}
\begin{pmatrix}
\conj{\alpha}-e^{-i\theta} & -\beta \\
\conj{\beta} & \alpha-e^{i\theta}
\end{pmatrix}.
\end{equation}

Although the spectrum of $\mathbb{U}$ is one dimensional, and hence depends only on a single real number $\theta$, we next convert the real integral into a contour integral. We define $\zeta=e^{i\theta}$ and, exploiting the fact that $\conj{\zeta}=\zeta^{-1}$ on the unit circle, write
\begin{equation}\label{eq:G_CUE_integral}
\mathcal{G}(Q)=\frac{1}{2\pi i}\oint_{\abs{\zeta}=1}\frac{\d\zeta}{\zeta}\frac{1}{(\alpha-\zeta)(\conj{\alpha}-\zeta^{-1})+\conj{\beta}\beta}
\begin{pmatrix}
\conj{\alpha}-\zeta^{-1} & -\beta \\
\conj{\beta} & \alpha-\zeta
\end{pmatrix}.
\end{equation}
Since the integrand is a holomorphic function of $\zeta$, the Green's function is the sum of the residues of the poles inside the unit circle.
The integrand has three poles,
\[
\zeta_0=0 
\qquad \text{and} \qquad \zeta_\pm=\frac{1+\conj{\alpha}\alpha+\conj{\beta}\beta\pm P}{2\conj{\alpha}},
\]
where $P=\sqrt{(1+\conj{\alpha}\alpha+\conj{\beta}\beta)^2-4\conj{\alpha}\alpha}$. Since $\zeta_-\zeta_+=\alpha/\conj{\alpha}$, i.e., $\abs{\zeta_-\zeta_+}=1$, one of $\zeta_\pm$ is inside the unit circle, the other outside. One can check that $\abs{\zeta_+}>\abs{\zeta_-}$, and hence conclude that $\zeta_0$ and $\zeta_-$ are always inside the unit circle and $\zeta_+$ outside. Computing the residues of the integrand of Eq.~(\ref{eq:G_CUE_integral}), we find
\begin{equation}\label{eq:G_CUE}
\mathcal{G}(Q)=\frac{1}{P}
\begin{pmatrix}
\frac{1}{2\alpha}(\conj{\alpha}\alpha-\conj{\beta}\beta-1+P) & -\beta \\
\conj{\beta} & \frac{1}{2\conj{\alpha}}(\conj{\alpha}\alpha-\conj{\beta}\beta-1+P)
\end{pmatrix}
\end{equation}

\subsection{Crossover ensemble}

\subsubsection{Quaternionic Resolvent}

We can now put the different pieces together. First, by recalling the scaling properties of $\mathcal{G}$ and $\mathcal{R}$, Eqs.~(\ref{eq:scaling_G})~and~(\ref{eq:scaling_B}), respectively, we have
\begin{equation}
\mathcal{R}_{aA}\left[
\begin{pmatrix}
\alpha & \beta \\
-\conj{\beta} & \conj{\alpha}
\end{pmatrix}
\right]=
a^2
\begin{pmatrix}
0 & \beta \\
-\conj{\beta} & 0
\end{pmatrix}
\end{equation}
and
\begin{equation}
\mathcal{G}_{bB}\left[
\begin{pmatrix}
\alpha & \beta \\
-\conj{\beta} & \conj{\alpha}
\end{pmatrix}
\right]=
\frac{1}{P_b}
\begin{pmatrix}
\frac{1}{2\alpha}(\conj{\alpha}\alpha-\conj{\beta}\beta-b^2+P_b) & -\beta \\
\conj{\beta} & \frac{1}{2\conj{\alpha}}(\conj{\alpha}\alpha-\conj{\beta}\beta-b^2+P_b)
\end{pmatrix},
\end{equation}
where $P_b=\sqrt{(b^2+\conj{\alpha}\alpha+\conj{\beta}\beta)^2-4b^2\conj{\alpha}\alpha}$. The non-Hermitian Pastur equation (\ref{eq:nonhermitian_Pastur}) then reads:
\begin{equation}
\begin{pmatrix}
\mathcal{G}_{11} & \mathcal{G}_{12}\\
-\conj{\mathcal{G}}_{12} & \conj{\mathcal{G}}_{11}
\end{pmatrix}=
\mathcal{G}_{bB}\left[
\begin{pmatrix}
z & -a^2\mathcal{G}_{12} \\
a^2\conj{\mathcal{G}}_{12} & \conj{z}
\end{pmatrix}
\right]=\frac{1}{P_b}
\begin{pmatrix}
\frac{1}{2z}(\conj{z}z-a^4\abs{\mathcal{G}_{12}}^2-b^2+P_b) & a^2\mathcal{G}_{12}\\
-a^2\conj{\mathcal{G}}_{12} & \frac{1}{2\conj{z}}(\conj{z}z-a^4\abs{\mathcal{G}_{12}}^2-b^2+P_b)
\end{pmatrix}.
\end{equation}


\subsubsection{Spectral Support}
The off-diagonal terms of the non-Hermitian Pastur equation yield the condition $\mathcal{G}_{12}=\mathcal{G}_{12}a^2/P_b$, which has two solutions. The trivial solution $\mathcal{G}_{12}=0$ is valid outside the eigenvalue support, while the nontrivial solution $P_b=a^2$ is satisfied inside. The value of $\mathcal{G}_{12}$ inside the support then satisfies
\begin{equation}\label{eq:G12_nontrivial}
a^4\abs{\mathcal{G}_{12}}^2=-b^2-\conj{z}z+\sqrt{a^4+4b^2\conj{z}z}.
\end{equation}

The boundaries of the spectral support are found by matching the trivial and nontrivial solutions, i.e., setting $\mathcal{G}_{12}=0$ in Eq.~(\ref{eq:G12_nontrivial}) and solving for $\conj{z}z$. The solutions match for $\abs{z}=\sqrt{b^2\pm a^2}$, and we therefore conclude  that the spectrum is supported on an annulus with inner (outer) radius $R_-(R_+)$, where $R_\pm=\sqrt{b^2\pm a^2}$. When $\abs{a}>\abs{b}$, there is no inner boundary and the spectrum is supported on the disk with radius $R_+$.

\subsubsection{Eigenvalue density}
The diagonal terms of the non-Hermitian Pastur equation (\ref{eq:nonhermitian_Pastur}) give the condition
\begin{equation}
\mathcal{G}_{11}=\frac{1}{2z}\left(1+\frac{\conj{z}z-a^4\abs{\mathcal{G}_{12}}^2-b^2}{P_b}\right).
\end{equation}
Using $P_b=a^2$ and Eq.~(\ref{eq:G12_nontrivial}), the value of $\mathcal{G}_{11}$ inside the spectral support is 
\begin{equation}
\mathcal{G}_{11}(z,\conj{z})=\frac{a^2+2\conj{z}z-\sqrt{a^4+4b^2\conj{z}z}}{2a^2z}.
\end{equation}

Finally, the eigenvalue distribution can be obtained by differentiating $\mathcal{G}_{11}$:
\begin{equation}
\varrho(z,\conj{z})=\frac{1}{\pi}\partial_{\conj{z}}\mathcal{G}_{11}(z,\conj{z})=\frac{1}{\pi a^2}\left(1-\frac{b^2}{\sqrt{a^4+4b^2\conj{z}z}}\right).
\end{equation}

By setting $a=p/\sqrt{d}$ and $b=(1-p)$, the expressions given in the main text are immediately recovered.

\subsection{Connection to the Single-Ring Theorem}
It is instructive to note that our results are in agreement with the single-ring theorem of Ref.~\cite{feinberg2001}. In particular, the single-ring theorem gives the inner and outer radii of the effective Kraus map from the moments
\begin{align}
    \left(R_{\pm} \right)^{\pm 2} = \frac{1}{N^{2}} \left\langle {\rm Tr} \left(\tilde\Phi^{\dagger} \tilde\Phi\right)^{\pm 1} \right\rangle.
\end{align}

Calculating these moments requires only the following result
\begin{align}
    {\rm lim}_{N\to \infty} \left\langle \frac{1}{N^{2}} {\rm Tr} \left( \mathbb{U}^{\dagger} \mathbb{G}\right)^{k} \left( \mathbb{G}^{\dagger} \mathbb{U}\right)^{k'} \right\rangle  = \delta_{kk'},
\end{align}
which follows by keeping only planar diagrams when Wick contracting over the $\mathbb{G}$ variables. Explicitly, for the outer radius we have the mean singular value of the effective Kraus map
\begin{align}
    R_{+}^{2} = (1 - p)^{2}\frac{1}{N^{2}} \left\langle {\rm Tr} \mathbb{U}^{\dagger} \mathbb{U}\right\rangle  + \frac{p^{2}}{d} \frac{1}{N^{2}} \left\langle {\rm Tr} \mathbb{G}^{\dagger} \mathbb{G}\right\rangle = (1 - p)^{2} +p^{2}/d,
\end{align}
whereas for the inner radius, we have
\begin{equation}
\begin{split}
    \frac{1}{R_{-}^{2}} &= \frac{1}{N^{2}} \left\langle {\rm Tr} \frac{1}{(1-p)^{2}} \left( 1 + \frac{p}{\sqrt{d}(1 - p)} \mathbb{U}^{\dagger} \mathbb{G}\right)^{-1} \left( 1 + \frac{p}{\sqrt{d}(1 - p)}  \mathbb{G}^{\dagger} \mathbb{U}\right)^{-1}\right\rangle\\
    & = \frac{1}{(1 - p)^{2}}\sum_{k}  \left( \frac{p^{2}}{d (1 - p)^{2}} \right)^{k} = \frac{1}{(1 - p)^{2} - p^{2}/d}.
\end{split}
\end{equation}

\twocolumngrid

\bibliography{bibfile}

\begin{thebibliography}{64}%
\makeatletter
\providecommand \@ifxundefined [1]{%
 \@ifx{#1\undefined}
}%
\providecommand \@ifnum [1]{%
 \ifnum #1\expandafter \@firstoftwo
 \else \expandafter \@secondoftwo
 \fi
}%
\providecommand \@ifx [1]{%
 \ifx #1\expandafter \@firstoftwo
 \else \expandafter \@secondoftwo
 \fi
}%
\providecommand \natexlab [1]{#1}%
\providecommand \enquote  [1]{``#1''}%
\providecommand \bibnamefont  [1]{#1}%
\providecommand \bibfnamefont [1]{#1}%
\providecommand \citenamefont [1]{#1}%
\providecommand \href@noop [0]{\@secondoftwo}%
\providecommand \href [0]{\begingroup \@sanitize@url \@href}%
\providecommand \@href[1]{\@@startlink{#1}\@@href}%
\providecommand \@@href[1]{\endgroup#1\@@endlink}%
\providecommand \@sanitize@url [0]{\catcode `\\12\catcode `\$12\catcode
  `\&12\catcode `\#12\catcode `\^12\catcode `\_12\catcode `\%12\relax}%
\providecommand \@@startlink[1]{}%
\providecommand \@@endlink[0]{}%
\providecommand \url  [0]{\begingroup\@sanitize@url \@url }%
\providecommand \@url [1]{\endgroup\@href {#1}{\urlprefix }}%
\providecommand \urlprefix  [0]{URL }%
\providecommand \Eprint [0]{\href }%
\providecommand \doibase [0]{https://doi.org/}%
\providecommand \selectlanguage [0]{\@gobble}%
\providecommand \bibinfo  [0]{\@secondoftwo}%
\providecommand \bibfield  [0]{\@secondoftwo}%
\providecommand \translation [1]{[#1]}%
\providecommand \BibitemOpen [0]{}%
\providecommand \bibitemStop [0]{}%
\providecommand \bibitemNoStop [0]{.\EOS\space}%
\providecommand \EOS [0]{\spacefactor3000\relax}%
\providecommand \BibitemShut  [1]{\csname bibitem#1\endcsname}%
\let\auto@bib@innerbib\@empty
\bibitem [{\citenamefont {Kraus}(1983)}]{kraus1983}%
  \BibitemOpen
  \bibfield  {author} {\bibinfo {author} {\bibfnamefont {K.}~\bibnamefont
  {Kraus}},\ }\href@noop {} {\emph {\bibinfo {title} {States, Effects and
  Operations: Fundamental Notions of Quantum Theory}}}\ (\bibinfo  {publisher}
  {Springer-Verlag},\ \bibinfo {address} {Berlin},\ \bibinfo {year}
  {1983})\BibitemShut {NoStop}%
\bibitem [{\citenamefont {Bohigas}\ \emph {et~al.}(1984)\citenamefont
  {Bohigas}, \citenamefont {Giannoni},\ and\ \citenamefont
  {Schmit}}]{bohigas1984}%
  \BibitemOpen
  \bibfield  {author} {\bibinfo {author} {\bibfnamefont {O.}~\bibnamefont
  {Bohigas}}, \bibinfo {author} {\bibfnamefont {M.-J.}\ \bibnamefont
  {Giannoni}},\ and\ \bibinfo {author} {\bibfnamefont {C.}~\bibnamefont
  {Schmit}},\ }\bibfield  {title} {\bibinfo {title} {Characterization of
  chaotic quantum spectra and universality of level fluctuation laws},\ }\href
  {http://dx.doi.org/10.1103/PhysRevLett.52.1} {\bibfield  {journal} {\bibinfo
  {journal} {Phys. Rev. Lett.}\ }\textbf {\bibinfo {volume} {52}},\ \bibinfo
  {pages} {1} (\bibinfo {year} {1984})}\BibitemShut {NoStop}%
\bibitem [{\citenamefont {Berry}\ and\ \citenamefont
  {Tabor}(1977)}]{berry1977}%
  \BibitemOpen
  \bibfield  {author} {\bibinfo {author} {\bibfnamefont {M.~V.}\ \bibnamefont
  {Berry}}\ and\ \bibinfo {author} {\bibfnamefont {M.}~\bibnamefont {Tabor}},\
  }\bibfield  {title} {\bibinfo {title} {Level clustering in the regular
  spectrum},\ }\href {https://doi.org/10.1098/rspa.1977.0140} {\bibfield
  {journal} {\bibinfo  {journal} {Proc. R. Soc. London, Ser. A}\ }\textbf
  {\bibinfo {volume} {356}},\ \bibinfo {pages} {375} (\bibinfo {year}
  {1977})}\BibitemShut {NoStop}%
\bibitem [{\citenamefont {Guhr}\ \emph {et~al.}(1998)\citenamefont {Guhr},
  \citenamefont {M{\"u}ller-Groeling},\ and\ \citenamefont
  {Weidenm{\"u}ller}}]{guhr1998}%
  \BibitemOpen
  \bibfield  {author} {\bibinfo {author} {\bibfnamefont {T.}~\bibnamefont
  {Guhr}}, \bibinfo {author} {\bibfnamefont {A.}~\bibnamefont
  {M{\"u}ller-Groeling}},\ and\ \bibinfo {author} {\bibfnamefont {H.~A.}\
  \bibnamefont {Weidenm{\"u}ller}},\ }\bibfield  {title} {\bibinfo {title}
  {Random-matrix theories in quantum physics: common concepts},\ }\href
  {https://doi.org/10.1016/S0370-1573(97)00088-4} {\bibfield  {journal}
  {\bibinfo  {journal} {Phys. Rep.}\ }\textbf {\bibinfo {volume} {299}},\
  \bibinfo {pages} {189} (\bibinfo {year} {1998})}\BibitemShut {NoStop}%
\bibitem [{\citenamefont {Beenakker}(1997)}]{beenaker1997}%
  \BibitemOpen
  \bibfield  {author} {\bibinfo {author} {\bibfnamefont {C.~W.~J.}\
  \bibnamefont {Beenakker}},\ }\bibfield  {title} {\bibinfo {title}
  {Random-matrix theory of quantum transport},\ }\href
  {https://doi.org/10.1103/RevModPhys.69.731} {\bibfield  {journal} {\bibinfo
  {journal} {Rev. Mod. Phys.}\ }\textbf {\bibinfo {volume} {69}},\ \bibinfo
  {pages} {731} (\bibinfo {year} {1997})}\BibitemShut {NoStop}%
\bibitem [{\citenamefont {Beenakker}(2015)}]{beenaker2015}%
  \BibitemOpen
  \bibfield  {author} {\bibinfo {author} {\bibfnamefont {C.~W.~J.}\
  \bibnamefont {Beenakker}},\ }\bibfield  {title} {\bibinfo {title}
  {Random-matrix theory of {Majorana} fermions and topological
  superconductors},\ }\href {https://doi.org/10.1103/RevModPhys.87.1037}
  {\bibfield  {journal} {\bibinfo  {journal} {Rev. Mod. Phys.}\ }\textbf
  {\bibinfo {volume} {87}},\ \bibinfo {pages} {1037} (\bibinfo {year}
  {2015})}\BibitemShut {NoStop}%
\bibitem [{\citenamefont {Schomerus}(2017)}]{schomerus2016}%
  \BibitemOpen
  \bibfield  {author} {\bibinfo {author} {\bibfnamefont {H.}~\bibnamefont
  {Schomerus}},\ }\bibfield  {title} {\bibinfo {title} {Random matrix
  approaches to open quantum systems},\ }in\ \href@noop {} {\emph {\bibinfo
  {booktitle} {Stochastic Processes and Random Matrices}}},\ \bibinfo {series}
  {Lecture Notes of the Les Houches Summer School}, Vol.\ \bibinfo {volume}
  {104}\ (\bibinfo  {publisher} {Oxford University Press},\ \bibinfo {address}
  {Oxford},\ \bibinfo {year} {2017})\ p.\ \bibinfo {pages} {409}\BibitemShut
  {NoStop}%
\bibitem [{\citenamefont {Sokolov}\ and\ \citenamefont
  {Zelevinsky}(1988)}]{sokolov1988}%
  \BibitemOpen
  \bibfield  {author} {\bibinfo {author} {\bibfnamefont {V.~V.}\ \bibnamefont
  {Sokolov}}\ and\ \bibinfo {author} {\bibfnamefont {V.~G.}\ \bibnamefont
  {Zelevinsky}},\ }\bibfield  {title} {\bibinfo {title} {On a statistical
  theory of overlapping resonances},\ }\href
  {https://doi.org/10.1016/0370-2693(88)90844-1} {\bibfield  {journal}
  {\bibinfo  {journal} {Phys. Lett. B}\ }\textbf {\bibinfo {volume} {202}},\
  \bibinfo {pages} {10} (\bibinfo {year} {1988})}\BibitemShut {NoStop}%
\bibitem [{\citenamefont {Sokolov}\ and\ \citenamefont
  {Zelevinsky}(1989)}]{sokolov1989}%
  \BibitemOpen
  \bibfield  {author} {\bibinfo {author} {\bibfnamefont {V.~V.}\ \bibnamefont
  {Sokolov}}\ and\ \bibinfo {author} {\bibfnamefont {V.~G.}\ \bibnamefont
  {Zelevinsky}},\ }\bibfield  {title} {\bibinfo {title} {Dynamics and
  statistics of unstable quantum states},\ }\href
  {https://doi.org/10.1016/0375-9474(89)90558-7} {\bibfield  {journal}
  {\bibinfo  {journal} {Nucl. Phys. A}\ }\textbf {\bibinfo {volume} {504}},\
  \bibinfo {pages} {562} (\bibinfo {year} {1989})}\BibitemShut {NoStop}%
\bibitem [{\citenamefont {Haake}\ \emph {et~al.}(1992)\citenamefont {Haake},
  \citenamefont {Izrailev}, \citenamefont {Lehmann}, \citenamefont {Saher},\
  and\ \citenamefont {Sommers}}]{haake1992}%
  \BibitemOpen
  \bibfield  {author} {\bibinfo {author} {\bibfnamefont {F.}~\bibnamefont
  {Haake}}, \bibinfo {author} {\bibfnamefont {F.}~\bibnamefont {Izrailev}},
  \bibinfo {author} {\bibfnamefont {N.}~\bibnamefont {Lehmann}}, \bibinfo
  {author} {\bibfnamefont {D.}~\bibnamefont {Saher}},\ and\ \bibinfo {author}
  {\bibfnamefont {H.-J.}\ \bibnamefont {Sommers}},\ }\bibfield  {title}
  {\bibinfo {title} {Statistics of complex levels of random matrices for
  decaying systems},\ }\href {https://doi.org/10.1007/BF01470925} {\bibfield
  {journal} {\bibinfo  {journal} {Z. Phys. B Condens. Matter}\ }\textbf
  {\bibinfo {volume} {88}},\ \bibinfo {pages} {359} (\bibinfo {year}
  {1992})}\BibitemShut {NoStop}%
\bibitem [{\citenamefont {Lehmann}\ \emph {et~al.}(1995)\citenamefont
  {Lehmann}, \citenamefont {Saher}, \citenamefont {Sokolov},\ and\
  \citenamefont {Sommers}}]{lehmann1995}%
  \BibitemOpen
  \bibfield  {author} {\bibinfo {author} {\bibfnamefont {N.}~\bibnamefont
  {Lehmann}}, \bibinfo {author} {\bibfnamefont {D.}~\bibnamefont {Saher}},
  \bibinfo {author} {\bibfnamefont {V.}~\bibnamefont {Sokolov}},\ and\ \bibinfo
  {author} {\bibfnamefont {H.-J.}\ \bibnamefont {Sommers}},\ }\bibfield
  {title} {\bibinfo {title} {Chaotic scattering: the supersymmetry method for
  large number of channels},\ }\href
  {https://doi.org/10.1016/0375-9474(94)00460-5} {\bibfield  {journal}
  {\bibinfo  {journal} {Nucl. Phys. A}\ }\textbf {\bibinfo {volume} {582}},\
  \bibinfo {pages} {223} (\bibinfo {year} {1995})}\BibitemShut {NoStop}%
\bibitem [{\citenamefont {Bruzda}\ \emph {et~al.}(2009)\citenamefont {Bruzda},
  \citenamefont {Cappellini}, \citenamefont {Sommers},\ and\ \citenamefont
  {{\.Z}yczkowski}}]{bruzda2009}%
  \BibitemOpen
  \bibfield  {author} {\bibinfo {author} {\bibfnamefont {W.}~\bibnamefont
  {Bruzda}}, \bibinfo {author} {\bibfnamefont {V.}~\bibnamefont {Cappellini}},
  \bibinfo {author} {\bibfnamefont {H.-J.}\ \bibnamefont {Sommers}},\ and\
  \bibinfo {author} {\bibfnamefont {K.}~\bibnamefont {{\.Z}yczkowski}},\
  }\bibfield  {title} {\bibinfo {title} {Random quantum operations},\ }\href
  {https://doi.org/10.1016/j.physleta.2008.11.043} {\bibfield  {journal}
  {\bibinfo  {journal} {Phys. Lett. A}\ }\textbf {\bibinfo {volume} {373}},\
  \bibinfo {pages} {320} (\bibinfo {year} {2009})}\BibitemShut {NoStop}%
\bibitem [{\citenamefont {Bruzda}\ \emph {et~al.}(2010)\citenamefont {Bruzda},
  \citenamefont {Smaczy{\'n}ski}, \citenamefont {Cappellini}, \citenamefont
  {Sommers},\ and\ \citenamefont {{\.Z}yczkowski}}]{bruzda2010}%
  \BibitemOpen
  \bibfield  {author} {\bibinfo {author} {\bibfnamefont {W.}~\bibnamefont
  {Bruzda}}, \bibinfo {author} {\bibfnamefont {M.}~\bibnamefont
  {Smaczy{\'n}ski}}, \bibinfo {author} {\bibfnamefont {V.}~\bibnamefont
  {Cappellini}}, \bibinfo {author} {\bibfnamefont {H.-J.}\ \bibnamefont
  {Sommers}},\ and\ \bibinfo {author} {\bibfnamefont {K.}~\bibnamefont
  {{\.Z}yczkowski}},\ }\bibfield  {title} {\bibinfo {title} {Universality of
  spectra for interacting quantum chaotic systems},\ }\href
  {https://doi.org/10.1103/PhysRevE.81.066209} {\bibfield  {journal} {\bibinfo
  {journal} {Phys. Rev. E}\ }\textbf {\bibinfo {volume} {81}},\ \bibinfo
  {pages} {066209} (\bibinfo {year} {2010})}\BibitemShut {NoStop}%
\bibitem [{\citenamefont {Gorin}\ and\ \citenamefont
  {Seligman}(2003)}]{gorin2003}%
  \BibitemOpen
  \bibfield  {author} {\bibinfo {author} {\bibfnamefont {T.}~\bibnamefont
  {Gorin}}\ and\ \bibinfo {author} {\bibfnamefont {T.~H.}\ \bibnamefont
  {Seligman}},\ }\bibfield  {title} {\bibinfo {title} {Decoherence in chaotic
  and integrable systems: a random matrix approach},\ }\href
  {https://doi.org/10.1016/S0375-9601(03)00131-2} {\bibfield  {journal}
  {\bibinfo  {journal} {Phys. Lett. A}\ }\textbf {\bibinfo {volume} {309}},\
  \bibinfo {pages} {61} (\bibinfo {year} {2003})}\BibitemShut {NoStop}%
\bibitem [{\citenamefont {Gorin}\ \emph {et~al.}(2008)\citenamefont {Gorin},
  \citenamefont {Pineda}, \citenamefont {Kohler},\ and\ \citenamefont
  {Seligman}}]{gorin2008}%
  \BibitemOpen
  \bibfield  {author} {\bibinfo {author} {\bibfnamefont {T.}~\bibnamefont
  {Gorin}}, \bibinfo {author} {\bibfnamefont {C.}~\bibnamefont {Pineda}},
  \bibinfo {author} {\bibfnamefont {H.}~\bibnamefont {Kohler}},\ and\ \bibinfo
  {author} {\bibfnamefont {T.}~\bibnamefont {Seligman}},\ }\bibfield  {title}
  {\bibinfo {title} {A random matrix theory of decoherence},\ }\href
  {https://doi.org/10.1088/1367-2630/10/11/115016} {\bibfield  {journal}
  {\bibinfo  {journal} {New J. Phys.}\ }\textbf {\bibinfo {volume} {10}},\
  \bibinfo {pages} {115016} (\bibinfo {year} {2008})}\BibitemShut {NoStop}%
\bibitem [{\citenamefont {Xu}\ \emph {et~al.}(2019)\citenamefont {Xu},
  \citenamefont {Garc\'{\i}a-Pintos}, \citenamefont {Chenu},\ and\
  \citenamefont {del Campo}}]{xu2019}%
  \BibitemOpen
  \bibfield  {author} {\bibinfo {author} {\bibfnamefont {Z.}~\bibnamefont
  {Xu}}, \bibinfo {author} {\bibfnamefont {L.~P.}\ \bibnamefont
  {Garc\'{\i}a-Pintos}}, \bibinfo {author} {\bibfnamefont {A.}~\bibnamefont
  {Chenu}},\ and\ \bibinfo {author} {\bibfnamefont {A.}~\bibnamefont {del
  Campo}},\ }\bibfield  {title} {\bibinfo {title} {Extreme decoherence and
  quantum chaos},\ }\href {https://doi.org/10.1103/PhysRevLett.122.014103}
  {\bibfield  {journal} {\bibinfo  {journal} {Phys. Rev. Lett.}\ }\textbf
  {\bibinfo {volume} {122}},\ \bibinfo {pages} {014103} (\bibinfo {year}
  {2019})}\BibitemShut {NoStop}%
\bibitem [{\citenamefont {Denisov}\ \emph {et~al.}(2019)\citenamefont
  {Denisov}, \citenamefont {Laptyeva}, \citenamefont {Tarnowski}, \citenamefont
  {Chru{\'s}ci{\'n}ski},\ and\ \citenamefont {{\.Z}yczkowski}}]{denisov2018}%
  \BibitemOpen
  \bibfield  {author} {\bibinfo {author} {\bibfnamefont {S.}~\bibnamefont
  {Denisov}}, \bibinfo {author} {\bibfnamefont {T.}~\bibnamefont {Laptyeva}},
  \bibinfo {author} {\bibfnamefont {W.}~\bibnamefont {Tarnowski}}, \bibinfo
  {author} {\bibfnamefont {D.}~\bibnamefont {Chru{\'s}ci{\'n}ski}},\ and\
  \bibinfo {author} {\bibfnamefont {K.}~\bibnamefont {{\.Z}yczkowski}},\
  }\bibfield  {title} {\bibinfo {title} {Universal spectra of random {Lindblad}
  operators},\ }\href {https://doi.org/10.1103/PhysRevLett.123.140403}
  {\bibfield  {journal} {\bibinfo  {journal} {Phys. Rev. Lett.}\ }\textbf
  {\bibinfo {volume} {123}},\ \bibinfo {pages} {140403} (\bibinfo {year}
  {2019})}\BibitemShut {NoStop}%
\bibitem [{\citenamefont {Can}\ \emph {et~al.}(2019)\citenamefont {Can},
  \citenamefont {Oganesyan}, \citenamefont {Orgad},\ and\ \citenamefont
  {Gopalakrishnan}}]{can2019prl}%
  \BibitemOpen
  \bibfield  {author} {\bibinfo {author} {\bibfnamefont {T.}~\bibnamefont
  {Can}}, \bibinfo {author} {\bibfnamefont {V.}~\bibnamefont {Oganesyan}},
  \bibinfo {author} {\bibfnamefont {D.}~\bibnamefont {Orgad}},\ and\ \bibinfo
  {author} {\bibfnamefont {S.}~\bibnamefont {Gopalakrishnan}},\ }\bibfield
  {title} {\bibinfo {title} {Spectral gaps and midgap states in random quantum
  master equations},\ }\href {https://doi.org/10.1103/PhysRevLett.123.234103}
  {\bibfield  {journal} {\bibinfo  {journal} {Phys. Rev. Lett.}\ }\textbf
  {\bibinfo {volume} {123}},\ \bibinfo {pages} {234103} (\bibinfo {year}
  {2019})}\BibitemShut {NoStop}%
\bibitem [{\citenamefont {Can}(2019)}]{can2019jphysa}%
  \BibitemOpen
  \bibfield  {author} {\bibinfo {author} {\bibfnamefont {T.}~\bibnamefont
  {Can}},\ }\bibfield  {title} {\bibinfo {title} {Random {Lindblad} dynamics},\
  }\href {https://doi.org/10.1088/1751-8121/ab4d26} {\bibfield  {journal}
  {\bibinfo  {journal} {J. Phys. A: Math. Theor.}\ }\textbf {\bibinfo {volume}
  {52}},\ \bibinfo {pages} {485302} (\bibinfo {year} {2019})}\BibitemShut
  {NoStop}%
\bibitem [{\citenamefont {S{\'a}}\ \emph {et~al.}(2020)\citenamefont {S{\'a}},
  \citenamefont {Ribeiro},\ and\ \citenamefont {Prosen}}]{sa2020RL}%
  \BibitemOpen
  \bibfield  {author} {\bibinfo {author} {\bibfnamefont {L.}~\bibnamefont
  {S{\'a}}}, \bibinfo {author} {\bibfnamefont {P.}~\bibnamefont {Ribeiro}},\
  and\ \bibinfo {author} {\bibfnamefont {T.}~\bibnamefont {Prosen}},\
  }\bibfield  {title} {\bibinfo {title} {Spectral and steady-state properties
  of random {Liouvillians}},\ }\href {https://doi.org/10.1088/1751-8121/ab9337}
  {\bibfield  {journal} {\bibinfo  {journal} {J. Phys. A: Math. Theor.}\
  }\textbf {\bibinfo {volume} {53}},\ \bibinfo {pages} {305303} (\bibinfo
  {year} {2020})}\BibitemShut {NoStop}%
\bibitem [{\citenamefont {Wang}\ \emph {et~al.}(2020)\citenamefont {Wang},
  \citenamefont {Piazza},\ and\ \citenamefont {Luitz}}]{wang2020}%
  \BibitemOpen
  \bibfield  {author} {\bibinfo {author} {\bibfnamefont {K.}~\bibnamefont
  {Wang}}, \bibinfo {author} {\bibfnamefont {F.}~\bibnamefont {Piazza}},\ and\
  \bibinfo {author} {\bibfnamefont {D.~J.}\ \bibnamefont {Luitz}},\ }\bibfield
  {title} {\bibinfo {title} {Hierarchy of relaxation timescales in local random
  {Liouvillians}},\ }\href {https://doi.org/10.1103/PhysRevLett.124.100604}
  {\bibfield  {journal} {\bibinfo  {journal} {Phys. Rev. Lett.}\ }\textbf
  {\bibinfo {volume} {124}},\ \bibinfo {pages} {100604} (\bibinfo {year}
  {2020})}\BibitemShut {NoStop}%
\bibitem [{\citenamefont {S\'a}\ \emph {et~al.}(2020)\citenamefont {S\'a},
  \citenamefont {Ribeiro},\ and\ \citenamefont {Prosen}}]{sa2019CSR}%
  \BibitemOpen
  \bibfield  {author} {\bibinfo {author} {\bibfnamefont {L.}~\bibnamefont
  {S\'a}}, \bibinfo {author} {\bibfnamefont {P.}~\bibnamefont {Ribeiro}},\ and\
  \bibinfo {author} {\bibfnamefont {T.}~\bibnamefont {Prosen}},\ }\bibfield
  {title} {\bibinfo {title} {Complex spacing ratios: A signature of dissipative
  quantum chaos},\ }\href {https://doi.org/10.1103/PhysRevX.10.021019}
  {\bibfield  {journal} {\bibinfo  {journal} {Phys. Rev. X}\ }\textbf {\bibinfo
  {volume} {10}},\ \bibinfo {pages} {021019} (\bibinfo {year}
  {2020})}\BibitemShut {NoStop}%
\bibitem [{\citenamefont {Akemann}\ \emph {et~al.}(2019)\citenamefont
  {Akemann}, \citenamefont {Kieburg}, \citenamefont {Mielke},\ and\
  \citenamefont {Prosen}}]{akemann2019}%
  \BibitemOpen
  \bibfield  {author} {\bibinfo {author} {\bibfnamefont {G.}~\bibnamefont
  {Akemann}}, \bibinfo {author} {\bibfnamefont {M.}~\bibnamefont {Kieburg}},
  \bibinfo {author} {\bibfnamefont {A.}~\bibnamefont {Mielke}},\ and\ \bibinfo
  {author} {\bibfnamefont {T.}~\bibnamefont {Prosen}},\ }\bibfield  {title}
  {\bibinfo {title} {Universal signature from integrability to chaos in
  dissipative open quantum systems},\ }\href
  {https://doi.org/10.1103/PhysRevLett.123.254101} {\bibfield  {journal}
  {\bibinfo  {journal} {Phys. Rev. Lett.}\ }\textbf {\bibinfo {volume} {123}},\
  \bibinfo {pages} {254101} (\bibinfo {year} {2019})}\BibitemShut {NoStop}%
\bibitem [{\citenamefont {Hamazaki}\ \emph {et~al.}(2020)\citenamefont
  {Hamazaki}, \citenamefont {Kawabata}, \citenamefont {Kura},\ and\
  \citenamefont {Ueda}}]{hamazaki2019}%
  \BibitemOpen
  \bibfield  {author} {\bibinfo {author} {\bibfnamefont {R.}~\bibnamefont
  {Hamazaki}}, \bibinfo {author} {\bibfnamefont {K.}~\bibnamefont {Kawabata}},
  \bibinfo {author} {\bibfnamefont {N.}~\bibnamefont {Kura}},\ and\ \bibinfo
  {author} {\bibfnamefont {M.}~\bibnamefont {Ueda}},\ }\bibfield  {title}
  {\bibinfo {title} {Universality classes of non-{Hermitian} random matrices},\
  }\href {https://doi.org/10.1103/PhysRevResearch.2.023286} {\bibfield
  {journal} {\bibinfo  {journal} {Phys. Rev. Research}\ }\textbf {\bibinfo
  {volume} {2}},\ \bibinfo {pages} {023286} (\bibinfo {year}
  {2020})}\BibitemShut {NoStop}%
\bibitem [{\citenamefont {Dyson}(1962)}]{dyson1962i}%
  \BibitemOpen
  \bibfield  {author} {\bibinfo {author} {\bibfnamefont {F.~J.}\ \bibnamefont
  {Dyson}},\ }\bibfield  {title} {\bibinfo {title} {Statistical theory of the
  energy levels of complex systems. {I}},\ }\href
  {https://doi.org/10.1063/1.1703773} {\bibfield  {journal} {\bibinfo
  {journal} {J. Math. Phys.}\ }\textbf {\bibinfo {volume} {3}},\ \bibinfo
  {pages} {140} (\bibinfo {year} {1962})}\BibitemShut {NoStop}%
\bibitem [{\citenamefont {Haake}(2013)}]{haake2013}%
  \BibitemOpen
  \bibfield  {author} {\bibinfo {author} {\bibfnamefont {F.}~\bibnamefont
  {Haake}},\ }\href@noop {} {\emph {\bibinfo {title} {Quantum signatures of
  chaos}}},\ Vol.~\bibinfo {volume} {54}\ (\bibinfo  {publisher} {Springer},\
  \bibinfo {address} {Berlin},\ \bibinfo {year} {2013})\BibitemShut {NoStop}%
\bibitem [{\citenamefont {Nielsen}\ and\ \citenamefont
  {Chuang}(2002)}]{nielsen2002}%
  \BibitemOpen
  \bibfield  {author} {\bibinfo {author} {\bibfnamefont {M.~A.}\ \bibnamefont
  {Nielsen}}\ and\ \bibinfo {author} {\bibfnamefont {I.}~\bibnamefont
  {Chuang}},\ }\href@noop {} {\emph {\bibinfo {title} {Quantum computation and
  quantum information}}}\ (\bibinfo  {publisher} {Cambridge University Press},\
  \bibinfo {address} {Cambridge},\ \bibinfo {year} {2002})\BibitemShut
  {NoStop}%
\bibitem [{\citenamefont {Bengtsson}\ and\ \citenamefont
  {{\.Z}yczkowski}(2017)}]{bengtsson2017}%
  \BibitemOpen
  \bibfield  {author} {\bibinfo {author} {\bibfnamefont {I.}~\bibnamefont
  {Bengtsson}}\ and\ \bibinfo {author} {\bibfnamefont {K.}~\bibnamefont
  {{\.Z}yczkowski}},\ }\href@noop {} {\emph {\bibinfo {title} {Geometry of
  quantum states: an introduction to quantum entanglement}}}\ (\bibinfo
  {publisher} {Cambridge university press},\ \bibinfo {address} {Cambridge},\
  \bibinfo {year} {2017})\BibitemShut {NoStop}%
\bibitem [{\citenamefont {{\.Z}yczkowski}\ and\ \citenamefont
  {Sommers}(2000)}]{zyczkowski2000}%
  \BibitemOpen
  \bibfield  {author} {\bibinfo {author} {\bibfnamefont {K.}~\bibnamefont
  {{\.Z}yczkowski}}\ and\ \bibinfo {author} {\bibfnamefont {H.-J.}\
  \bibnamefont {Sommers}},\ }\bibfield  {title} {\bibinfo {title} {Truncations
  of random unitary matrices},\ }\href
  {https://doi.org/10.1088/0305-4470/33/10/307} {\bibfield  {journal} {\bibinfo
   {journal} {J. Phys. A: Math. Gen.}\ }\textbf {\bibinfo {volume} {33}},\
  \bibinfo {pages} {2045} (\bibinfo {year} {2000})}\BibitemShut {NoStop}%
\bibitem [{\citenamefont {Ginibre}(1965)}]{ginibre1965}%
  \BibitemOpen
  \bibfield  {author} {\bibinfo {author} {\bibfnamefont {J.}~\bibnamefont
  {Ginibre}},\ }\bibfield  {title} {\bibinfo {title} {Statistical ensembles of
  complex, quaternion, and real matrices},\ }\href
  {https://doi.org/10.1063/1.1704292} {\bibfield  {journal} {\bibinfo
  {journal} {J. Math. Phys.}\ }\textbf {\bibinfo {volume} {6}},\ \bibinfo
  {pages} {440} (\bibinfo {year} {1965})}\BibitemShut {NoStop}%
\bibitem [{\citenamefont {Nica}\ and\ \citenamefont
  {Speicher}(2006)}]{nica2006}%
  \BibitemOpen
  \bibfield  {author} {\bibinfo {author} {\bibfnamefont {A.}~\bibnamefont
  {Nica}}\ and\ \bibinfo {author} {\bibfnamefont {R.}~\bibnamefont
  {Speicher}},\ }\href@noop {} {\emph {\bibinfo {title} {Lectures on the
  combinatorics of free probability}}},\ Vol.~\bibinfo {volume} {13}\ (\bibinfo
   {publisher} {Cambridge University Press},\ \bibinfo {address} {Cambridge},\
  \bibinfo {year} {2006})\BibitemShut {NoStop}%
\bibitem [{\citenamefont {Mingo}\ and\ \citenamefont
  {Speicher}(2017)}]{mingo2017}%
  \BibitemOpen
  \bibfield  {author} {\bibinfo {author} {\bibfnamefont {J.~A.}\ \bibnamefont
  {Mingo}}\ and\ \bibinfo {author} {\bibfnamefont {R.}~\bibnamefont
  {Speicher}},\ }\href@noop {} {\emph {\bibinfo {title} {Free probability and
  random matrices}}},\ Vol.~\bibinfo {volume} {35}\ (\bibinfo  {publisher}
  {Springer},\ \bibinfo {address} {New York},\ \bibinfo {year}
  {2017})\BibitemShut {NoStop}%
\bibitem [{\citenamefont {Mar{\v{c}}enko}\ and\ \citenamefont
  {Pastur}(1967)}]{marchenko1967}%
  \BibitemOpen
  \bibfield  {author} {\bibinfo {author} {\bibfnamefont {V.~A.}\ \bibnamefont
  {Mar{\v{c}}enko}}\ and\ \bibinfo {author} {\bibfnamefont {L.~A.}\
  \bibnamefont {Pastur}},\ }\bibfield  {title} {\bibinfo {title} {Distribution
  of eigenvalues for some sets of random matrices},\ }\href
  {https://doi.org/10.1070/SM1967v001n04ABEH001994} {\bibfield  {journal}
  {\bibinfo  {journal} {Math. USSR Sb.}\ }\textbf {\bibinfo {volume} {1}},\
  \bibinfo {pages} {457} (\bibinfo {year} {1967})}\BibitemShut {NoStop}%
\bibitem [{\citenamefont {{\.Z}yczkowski}\ and\ \citenamefont
  {Sommers}(2001)}]{zyczkowski2001}%
  \BibitemOpen
  \bibfield  {author} {\bibinfo {author} {\bibfnamefont {K.}~\bibnamefont
  {{\.Z}yczkowski}}\ and\ \bibinfo {author} {\bibfnamefont {H.-J.}\
  \bibnamefont {Sommers}},\ }\bibfield  {title} {\bibinfo {title} {Induced
  measures in the space of mixed quantum states},\ }\href
  {https://doi.org/10.1088/0305-4470/34/35/335} {\bibfield  {journal} {\bibinfo
   {journal} {J. Phys. A: Math. Gen.}\ }\textbf {\bibinfo {volume} {34}},\
  \bibinfo {pages} {7111} (\bibinfo {year} {2001})}\BibitemShut {NoStop}%
\bibitem [{\citenamefont {Sommers}\ and\ \citenamefont
  {{\.Z}yczkowski}(2004)}]{sommers2004}%
  \BibitemOpen
  \bibfield  {author} {\bibinfo {author} {\bibfnamefont {H.-J.}\ \bibnamefont
  {Sommers}}\ and\ \bibinfo {author} {\bibfnamefont {K.}~\bibnamefont
  {{\.Z}yczkowski}},\ }\bibfield  {title} {\bibinfo {title} {Statistical
  properties of random density matrices},\ }\href
  {https://doi.org/10.1088/0305-4470/37/35/004} {\bibfield  {journal} {\bibinfo
   {journal} {J. Phys. A: Math. Gen.}\ }\textbf {\bibinfo {volume} {37}},\
  \bibinfo {pages} {8457} (\bibinfo {year} {2004})}\BibitemShut {NoStop}%
\bibitem [{\citenamefont {{\v{Z}}nidari{\v{c}}}(2006)}]{znidaric2006}%
  \BibitemOpen
  \bibfield  {author} {\bibinfo {author} {\bibfnamefont {M.}~\bibnamefont
  {{\v{Z}}nidari{\v{c}}}},\ }\bibfield  {title} {\bibinfo {title} {Entanglement
  of random vectors},\ }\href {https://doi.org/10.1088/1751-8113/40/3/F04}
  {\bibfield  {journal} {\bibinfo  {journal} {J. Phys. A: Math. Theor.}\
  }\textbf {\bibinfo {volume} {40}},\ \bibinfo {pages} {F105} (\bibinfo {year}
  {2006})}\BibitemShut {NoStop}%
\bibitem [{\citenamefont {{\.Z}yczkowski}\ \emph {et~al.}(2011)\citenamefont
  {{\.Z}yczkowski}, \citenamefont {Penson}, \citenamefont {Nechita},\ and\
  \citenamefont {Collins}}]{zyczkowski2011}%
  \BibitemOpen
  \bibfield  {author} {\bibinfo {author} {\bibfnamefont {K.}~\bibnamefont
  {{\.Z}yczkowski}}, \bibinfo {author} {\bibfnamefont {K.~A.}\ \bibnamefont
  {Penson}}, \bibinfo {author} {\bibfnamefont {I.}~\bibnamefont {Nechita}},\
  and\ \bibinfo {author} {\bibfnamefont {B.}~\bibnamefont {Collins}},\
  }\bibfield  {title} {\bibinfo {title} {Generating random density matrices},\
  }\href {https://doi.org/10.1063/1.3595693} {\bibfield  {journal} {\bibinfo
  {journal} {J. Math. Phys.}\ }\textbf {\bibinfo {volume} {52}},\ \bibinfo
  {pages} {062201} (\bibinfo {year} {2011})}\BibitemShut {NoStop}%
\bibitem [{\citenamefont {Yang}\ \emph {et~al.}(2015)\citenamefont {Yang},
  \citenamefont {Chamon}, \citenamefont {Hamma},\ and\ \citenamefont
  {Mucciolo}}]{yang2015}%
  \BibitemOpen
  \bibfield  {author} {\bibinfo {author} {\bibfnamefont {Z.-C.}\ \bibnamefont
  {Yang}}, \bibinfo {author} {\bibfnamefont {C.}~\bibnamefont {Chamon}},
  \bibinfo {author} {\bibfnamefont {A.}~\bibnamefont {Hamma}},\ and\ \bibinfo
  {author} {\bibfnamefont {E.~R.}\ \bibnamefont {Mucciolo}},\ }\bibfield
  {title} {\bibinfo {title} {Two-component structure in the entanglement
  spectrum of highly excited states},\ }\href
  {https://doi.org/10.1103/PhysRevLett.115.267206} {\bibfield  {journal}
  {\bibinfo  {journal} {Phys. Rev. Lett.}\ }\textbf {\bibinfo {volume} {115}},\
  \bibinfo {pages} {267206} (\bibinfo {year} {2015})}\BibitemShut {NoStop}%
\bibitem [{\citenamefont {Forrester}(2010)}]{forrester2010}%
  \BibitemOpen
  \bibfield  {author} {\bibinfo {author} {\bibfnamefont {P.~J.}\ \bibnamefont
  {Forrester}},\ }\href@noop {} {\emph {\bibinfo {title} {Log-gases and random
  matrices}}}\ (\bibinfo  {publisher} {Princeton University Press},\ \bibinfo
  {address} {Princeton},\ \bibinfo {year} {2010})\BibitemShut {NoStop}%
\bibitem [{\citenamefont {Nadal}\ \emph {et~al.}(2011)\citenamefont {Nadal},
  \citenamefont {Majumdar},\ and\ \citenamefont {Vergassola}}]{nadal2011}%
  \BibitemOpen
  \bibfield  {author} {\bibinfo {author} {\bibfnamefont {C.}~\bibnamefont
  {Nadal}}, \bibinfo {author} {\bibfnamefont {S.~N.}\ \bibnamefont
  {Majumdar}},\ and\ \bibinfo {author} {\bibfnamefont {M.}~\bibnamefont
  {Vergassola}},\ }\bibfield  {title} {\bibinfo {title} {Statistical
  distribution of quantum entanglement for a random bipartite state},\ }\href
  {https://doi.org/10.1007/s10955-010-0108-4} {\bibfield  {journal} {\bibinfo
  {journal} {J. Stat. Phys.}\ }\textbf {\bibinfo {volume} {142}},\ \bibinfo
  {pages} {403} (\bibinfo {year} {2011})}\BibitemShut {NoStop}%
\bibitem [{\citenamefont {{\.Z}yczkowski}\ \emph {et~al.}(2003)\citenamefont
  {{\.Z}yczkowski}, \citenamefont {Kus}, \citenamefont {S{\l}omczy{\'n}ski},\
  and\ \citenamefont {Sommers}}]{zyczkowski2003}%
  \BibitemOpen
  \bibfield  {author} {\bibinfo {author} {\bibfnamefont {K.}~\bibnamefont
  {{\.Z}yczkowski}}, \bibinfo {author} {\bibfnamefont {M.}~\bibnamefont {Kus}},
  \bibinfo {author} {\bibfnamefont {W.}~\bibnamefont {S{\l}omczy{\'n}ski}},\
  and\ \bibinfo {author} {\bibfnamefont {H.-J.}\ \bibnamefont {Sommers}},\
  }\bibfield  {title} {\bibinfo {title} {Random unistochastic matrices},\
  }\href {https://doi.org/10.1088/0305-4470/36/12/333} {\bibfield  {journal}
  {\bibinfo  {journal} {J. Phys. A: Math. Gen.}\ }\textbf {\bibinfo {volume}
  {36}},\ \bibinfo {pages} {3425} (\bibinfo {year} {2003})}\BibitemShut
  {NoStop}%
\bibitem [{\citenamefont {Horvat}(2009)}]{horvat2009}%
  \BibitemOpen
  \bibfield  {author} {\bibinfo {author} {\bibfnamefont {M.}~\bibnamefont
  {Horvat}},\ }\bibfield  {title} {\bibinfo {title} {The ensemble of random
  {Markov} matrices},\ }\href
  {https://doi.org/10.1088/1742-5468/2009/07/P07005} {\bibfield  {journal}
  {\bibinfo  {journal} {J. Stat. Mech.: Theory Exp.}\ }\textbf {\bibinfo
  {volume} {2009}}\bibinfo  {number} { (07)},\ \bibinfo {pages}
  {P07005}}\BibitemShut {NoStop}%
\bibitem [{\citenamefont {Chafa{\"\i}}(2010)}]{chafai2010}%
  \BibitemOpen
\bibfield  {number} {  }\bibfield  {author} {\bibinfo {author} {\bibfnamefont
  {D.}~\bibnamefont {Chafa{\"\i}}},\ }\bibfield  {title} {\bibinfo {title} {The
  {Dirichlet Markov} ensemble},\ }\href
  {https://doi.org/10.1016/j.jmva.2009.10.013} {\bibfield  {journal} {\bibinfo
  {journal} {J. Multivariate Anal.}\ }\textbf {\bibinfo {volume} {101}},\
  \bibinfo {pages} {555} (\bibinfo {year} {2010})}\BibitemShut {NoStop}%
\bibitem [{\citenamefont {Pereyra}\ and\ \citenamefont
  {Mello}(1983)}]{pereyra1983}%
  \BibitemOpen
  \bibfield  {author} {\bibinfo {author} {\bibfnamefont {P.}~\bibnamefont
  {Pereyra}}\ and\ \bibinfo {author} {\bibfnamefont {P.}~\bibnamefont
  {Mello}},\ }\bibfield  {title} {\bibinfo {title} {Marginal distribution of
  the s-matrix elements for {Dyson's} measure and some applications},\ }\href
  {https://doi.org/10.1088/0305-4470/16/2/007} {\bibfield  {journal} {\bibinfo
  {journal} {J. Phys. A: Math. Gen.}\ }\textbf {\bibinfo {volume} {16}},\
  \bibinfo {pages} {237} (\bibinfo {year} {1983})}\BibitemShut {NoStop}%
\bibitem [{\citenamefont {Timm}(2009)}]{timm2009}%
  \BibitemOpen
  \bibfield  {author} {\bibinfo {author} {\bibfnamefont {C.}~\bibnamefont
  {Timm}},\ }\bibfield  {title} {\bibinfo {title} {Random transition-rate
  matrices for the master equation},\ }\href
  {https://doi.org/10.1103/PhysRevE.80.021140} {\bibfield  {journal} {\bibinfo
  {journal} {Phys. Rev. E}\ }\textbf {\bibinfo {volume} {80}},\ \bibinfo
  {pages} {021140} (\bibinfo {year} {2009})}\BibitemShut {NoStop}%
\bibitem [{\citenamefont {Srivastava}\ \emph {et~al.}(2018)\citenamefont
  {Srivastava}, \citenamefont {Lakshminarayan}, \citenamefont {Tomsovic},\ and\
  \citenamefont {B{\"a}cker}}]{srivastava2018}%
  \BibitemOpen
  \bibfield  {author} {\bibinfo {author} {\bibfnamefont {S.~C.}\ \bibnamefont
  {Srivastava}}, \bibinfo {author} {\bibfnamefont {A.}~\bibnamefont
  {Lakshminarayan}}, \bibinfo {author} {\bibfnamefont {S.}~\bibnamefont
  {Tomsovic}},\ and\ \bibinfo {author} {\bibfnamefont {A.}~\bibnamefont
  {B{\"a}cker}},\ }\bibfield  {title} {\bibinfo {title} {Ordered level spacing
  probability densities},\ }\href {https://doi.org/10.1088/1751-8121/aaefa4}
  {\bibfield  {journal} {\bibinfo  {journal} {J. Phys. A: Math. Theor.}\
  }\textbf {\bibinfo {volume} {52}},\ \bibinfo {pages} {025101} (\bibinfo
  {year} {2018})}\BibitemShut {NoStop}%
\bibitem [{\citenamefont {Oganesyan}\ and\ \citenamefont
  {Huse}(2007)}]{oganesyan2007}%
  \BibitemOpen
  \bibfield  {author} {\bibinfo {author} {\bibfnamefont {V.}~\bibnamefont
  {Oganesyan}}\ and\ \bibinfo {author} {\bibfnamefont {D.~A.}\ \bibnamefont
  {Huse}},\ }\bibfield  {title} {\bibinfo {title} {Localization of interacting
  fermions at high temperature},\ }\href
  {https://doi.org/10.1103/PhysRevB.75.155111} {\bibfield  {journal} {\bibinfo
  {journal} {Phys. Rev. B}\ }\textbf {\bibinfo {volume} {75}},\ \bibinfo
  {pages} {155111} (\bibinfo {year} {2007})}\BibitemShut {NoStop}%
\bibitem [{\citenamefont {Atas}\ \emph
  {et~al.}(2013{\natexlab{a}})\citenamefont {Atas}, \citenamefont {Bogomolny},
  \citenamefont {Giraud},\ and\ \citenamefont {Roux}}]{atas2013}%
  \BibitemOpen
  \bibfield  {author} {\bibinfo {author} {\bibfnamefont {Y.}~\bibnamefont
  {Atas}}, \bibinfo {author} {\bibfnamefont {E.}~\bibnamefont {Bogomolny}},
  \bibinfo {author} {\bibfnamefont {O.}~\bibnamefont {Giraud}},\ and\ \bibinfo
  {author} {\bibfnamefont {G.}~\bibnamefont {Roux}},\ }\bibfield  {title}
  {\bibinfo {title} {Distribution of the ratio of consecutive level spacings in
  random matrix ensembles},\ }\href
  {http://dx.doi.org/10.1103/PhysRevLett.110.084101} {\bibfield  {journal}
  {\bibinfo  {journal} {Phys. Rev. Lett.}\ }\textbf {\bibinfo {volume} {110}},\
  \bibinfo {pages} {084101} (\bibinfo {year} {2013}{\natexlab{a}})}\BibitemShut
  {NoStop}%
\bibitem [{\citenamefont {Atas}\ \emph
  {et~al.}(2013{\natexlab{b}})\citenamefont {Atas}, \citenamefont {Bogomolny},
  \citenamefont {Giraud}, \citenamefont {Vivo},\ and\ \citenamefont
  {Vivo}}]{atas2013long}%
  \BibitemOpen
  \bibfield  {author} {\bibinfo {author} {\bibfnamefont {Y.}~\bibnamefont
  {Atas}}, \bibinfo {author} {\bibfnamefont {E.}~\bibnamefont {Bogomolny}},
  \bibinfo {author} {\bibfnamefont {O.}~\bibnamefont {Giraud}}, \bibinfo
  {author} {\bibfnamefont {P.}~\bibnamefont {Vivo}},\ and\ \bibinfo {author}
  {\bibfnamefont {E.}~\bibnamefont {Vivo}},\ }\bibfield  {title} {\bibinfo
  {title} {Joint probability densities of level spacing ratios in random
  matrices},\ }\href {https://doi.org/10.1088/1751-8113/46/35/355204}
  {\bibfield  {journal} {\bibinfo  {journal} {J. Phys. A: Math. Theor.}\
  }\textbf {\bibinfo {volume} {46}},\ \bibinfo {pages} {355204} (\bibinfo
  {year} {2013}{\natexlab{b}})}\BibitemShut {NoStop}%
\bibitem [{\citenamefont {Forrester}\ and\ \citenamefont
  {Nagao}(2007)}]{forrester2007}%
  \BibitemOpen
  \bibfield  {author} {\bibinfo {author} {\bibfnamefont {P.~J.}\ \bibnamefont
  {Forrester}}\ and\ \bibinfo {author} {\bibfnamefont {T.}~\bibnamefont
  {Nagao}},\ }\bibfield  {title} {\bibinfo {title} {Eigenvalue statistics of
  the real {Ginibre} ensemble},\ }\href
  {https://doi.org/10.1103/PhysRevLett.99.050603} {\bibfield  {journal}
  {\bibinfo  {journal} {Phys. Rev. Lett.}\ }\textbf {\bibinfo {volume} {99}},\
  \bibinfo {pages} {050603} (\bibinfo {year} {2007})}\BibitemShut {NoStop}%
\bibitem [{\citenamefont {Feinberg}\ and\ \citenamefont
  {Zee}(1997{\natexlab{a}})}]{feinberg1997a}%
  \BibitemOpen
  \bibfield  {author} {\bibinfo {author} {\bibfnamefont {J.}~\bibnamefont
  {Feinberg}}\ and\ \bibinfo {author} {\bibfnamefont {A.}~\bibnamefont {Zee}},\
  }\bibfield  {title} {\bibinfo {title} {Non-hermitian random matrix theory:
  Method of hermitian reduction},\ }\href
  {https://doi.org/10.1016/S0550-3213(97)00502-6} {\bibfield  {journal}
  {\bibinfo  {journal} {Nucl. Phys. B}\ }\textbf {\bibinfo {volume} {504}},\
  \bibinfo {pages} {579} (\bibinfo {year} {1997}{\natexlab{a}})}\BibitemShut
  {NoStop}%
\bibitem [{\citenamefont {Feinberg}\ and\ \citenamefont
  {Zee}(1997{\natexlab{b}})}]{feinberg1997b}%
  \BibitemOpen
  \bibfield  {author} {\bibinfo {author} {\bibfnamefont {J.}~\bibnamefont
  {Feinberg}}\ and\ \bibinfo {author} {\bibfnamefont {A.}~\bibnamefont {Zee}},\
  }\bibfield  {title} {\bibinfo {title} {Non-{Gaussian} non-{Hermitian} random
  matrix theory: phase transition and addition formalism},\ }\href
  {https://doi.org/10.1016/S0550-3213(97)00419-7} {\bibfield  {journal}
  {\bibinfo  {journal} {Nucl. Phys. B}\ }\textbf {\bibinfo {volume} {501}},\
  \bibinfo {pages} {643} (\bibinfo {year} {1997}{\natexlab{b}})}\BibitemShut
  {NoStop}%
\bibitem [{\citenamefont {Janik}\ \emph
  {et~al.}(1997{\natexlab{a}})\citenamefont {Janik}, \citenamefont {Nowak},
  \citenamefont {Papp},\ and\ \citenamefont {Zahed}}]{janik1997a}%
  \BibitemOpen
  \bibfield  {author} {\bibinfo {author} {\bibfnamefont {R.~A.}\ \bibnamefont
  {Janik}}, \bibinfo {author} {\bibfnamefont {M.~A.}\ \bibnamefont {Nowak}},
  \bibinfo {author} {\bibfnamefont {G.}~\bibnamefont {Papp}},\ and\ \bibinfo
  {author} {\bibfnamefont {I.}~\bibnamefont {Zahed}},\ }\bibfield  {title}
  {\bibinfo {title} {Non-{Hermitian} random matrix models},\ }\href
  {https://doi.org/10.1016/S0550-3213(97)00418-5} {\bibfield  {journal}
  {\bibinfo  {journal} {Nucl. Phys. B}\ }\textbf {\bibinfo {volume} {501}},\
  \bibinfo {pages} {603} (\bibinfo {year} {1997}{\natexlab{a}})}\BibitemShut
  {NoStop}%
\bibitem [{\citenamefont {Janik}\ \emph
  {et~al.}(1997{\natexlab{b}})\citenamefont {Janik}, \citenamefont {Nowak},
  \citenamefont {Papp}, \citenamefont {Wambach},\ and\ \citenamefont
  {Zahed}}]{janik1997b}%
  \BibitemOpen
  \bibfield  {author} {\bibinfo {author} {\bibfnamefont {R.~A.}\ \bibnamefont
  {Janik}}, \bibinfo {author} {\bibfnamefont {M.~A.}\ \bibnamefont {Nowak}},
  \bibinfo {author} {\bibfnamefont {G.}~\bibnamefont {Papp}}, \bibinfo {author}
  {\bibfnamefont {J.}~\bibnamefont {Wambach}},\ and\ \bibinfo {author}
  {\bibfnamefont {I.}~\bibnamefont {Zahed}},\ }\bibfield  {title} {\bibinfo
  {title} {Non-{Hermitian} random matrix models: Free random variable
  approach},\ }\href {https://doi.org/10.1103/PhysRevE.55.4100} {\bibfield
  {journal} {\bibinfo  {journal} {Phys. Rev. E}\ }\textbf {\bibinfo {volume}
  {55}},\ \bibinfo {pages} {4100} (\bibinfo {year}
  {1997}{\natexlab{b}})}\BibitemShut {NoStop}%
\bibitem [{\citenamefont {Feinberg}\ \emph {et~al.}(2001)\citenamefont
  {Feinberg}, \citenamefont {Scalettar},\ and\ \citenamefont
  {Zee}}]{feinberg2001}%
  \BibitemOpen
  \bibfield  {author} {\bibinfo {author} {\bibfnamefont {J.}~\bibnamefont
  {Feinberg}}, \bibinfo {author} {\bibfnamefont {R.}~\bibnamefont
  {Scalettar}},\ and\ \bibinfo {author} {\bibfnamefont {A.}~\bibnamefont
  {Zee}},\ }\bibfield  {title} {\bibinfo {title} {``{Single} ring theorem'' and
  the disk-annulus phase transition},\ }\href
  {https://doi.org/10.1063/1.1412599} {\bibfield  {journal} {\bibinfo
  {journal} {J. Math. Phys.}\ }\textbf {\bibinfo {volume} {42}},\ \bibinfo
  {pages} {5718} (\bibinfo {year} {2001})}\BibitemShut {NoStop}%
\bibitem [{\citenamefont {Janik}\ \emph {et~al.}(2001)\citenamefont {Janik},
  \citenamefont {Nowak}, \citenamefont {Papp},\ and\ \citenamefont
  {Zahed}}]{janik2001}%
  \BibitemOpen
  \bibfield  {author} {\bibinfo {author} {\bibfnamefont {R.~A.}\ \bibnamefont
  {Janik}}, \bibinfo {author} {\bibfnamefont {M.~A.}\ \bibnamefont {Nowak}},
  \bibinfo {author} {\bibfnamefont {G.}~\bibnamefont {Papp}},\ and\ \bibinfo
  {author} {\bibfnamefont {I.}~\bibnamefont {Zahed}},\ }\bibfield  {title}
  {\bibinfo {title} {Green's functions in non-hermitian random matrix models},\
  }\href {https://doi.org/10.1016/S1386-9477(00)00244-7} {\bibfield  {journal}
  {\bibinfo  {journal} {Physica E}\ }\textbf {\bibinfo {volume} {9}},\ \bibinfo
  {pages} {456} (\bibinfo {year} {2001})}\BibitemShut {NoStop}%
\bibitem [{\citenamefont {Jarosz}\ and\ \citenamefont
  {Nowak}(2004)}]{jarosz2004}%
  \BibitemOpen
  \bibfield  {author} {\bibinfo {author} {\bibfnamefont {A.}~\bibnamefont
  {Jarosz}}\ and\ \bibinfo {author} {\bibfnamefont {M.~A.}\ \bibnamefont
  {Nowak}},\ }\bibfield  {title} {\bibinfo {title} {A novel approach to
  non-{Hermitian} random matrix models},\ }\href
  {https://arxiv.org/abs/math-ph/0402057} {\bibfield  {journal} {\bibinfo
  {journal} {arXiv:math-ph/0402057}\ } (\bibinfo {year} {2004})}\BibitemShut
  {NoStop}%
\bibitem [{\citenamefont {Jarosz}\ and\ \citenamefont
  {Nowak}(2006)}]{jarosz2006}%
  \BibitemOpen
  \bibfield  {author} {\bibinfo {author} {\bibfnamefont {A.}~\bibnamefont
  {Jarosz}}\ and\ \bibinfo {author} {\bibfnamefont {M.~A.}\ \bibnamefont
  {Nowak}},\ }\bibfield  {title} {\bibinfo {title} {Random {Hermitian} versus
  random non-{Hermitian} operators---unexpected links},\ }\href
  {https://doi.org/10.1088/0305-4470/39/32/S12} {\bibfield  {journal} {\bibinfo
   {journal} {J. Phys. A: Math. Gen.}\ }\textbf {\bibinfo {volume} {39}},\
  \bibinfo {pages} {10107} (\bibinfo {year} {2006})}\BibitemShut {NoStop}%
\bibitem [{\citenamefont {Feinberg}(2006)}]{feinberg2006}%
  \BibitemOpen
  \bibfield  {author} {\bibinfo {author} {\bibfnamefont {J.}~\bibnamefont
  {Feinberg}},\ }\bibfield  {title} {\bibinfo {title} {Non-{Hermitian} random
  matrix theory: summation of planar diagrams, the `single-ring' theorem and
  the disk--annulus phase transition},\ }\href
  {https://doi.org/10.1088/0305-4470/39/32/S07} {\bibfield  {journal} {\bibinfo
   {journal} {J. Phys. A: Math. Gen.}\ }\textbf {\bibinfo {volume} {39}},\
  \bibinfo {pages} {10029} (\bibinfo {year} {2006})}\BibitemShut {NoStop}%
\bibitem [{\citenamefont {Burda}\ \emph {et~al.}(2011)\citenamefont {Burda},
  \citenamefont {Janik},\ and\ \citenamefont {Nowak}}]{burda2011}%
  \BibitemOpen
  \bibfield  {author} {\bibinfo {author} {\bibfnamefont {Z.}~\bibnamefont
  {Burda}}, \bibinfo {author} {\bibfnamefont {R.~A.}\ \bibnamefont {Janik}},\
  and\ \bibinfo {author} {\bibfnamefont {M.~A.}\ \bibnamefont {Nowak}},\
  }\bibfield  {title} {\bibinfo {title} {Multiplication law and $s$ transform
  for {non-Hermitian} random matrices},\ }\href
  {https://doi.org/10.1103/PhysRevE.84.061125} {\bibfield  {journal} {\bibinfo
  {journal} {Phys. Rev. E}\ }\textbf {\bibinfo {volume} {84}},\ \bibinfo
  {pages} {061125} (\bibinfo {year} {2011})}\BibitemShut {NoStop}%
\bibitem [{\citenamefont {Burda}\ and\ \citenamefont
  {Swiech}(2015)}]{burda2015}%
  \BibitemOpen
  \bibfield  {author} {\bibinfo {author} {\bibfnamefont {Z.}~\bibnamefont
  {Burda}}\ and\ \bibinfo {author} {\bibfnamefont {A.}~\bibnamefont {Swiech}},\
  }\bibfield  {title} {\bibinfo {title} {Quaternionic {$R$} transform and
  {non-Hermitian} random matrices},\ }\href
  {https://doi.org/10.1103/PhysRevE.92.052111} {\bibfield  {journal} {\bibinfo
  {journal} {Phys. Rev. E}\ }\textbf {\bibinfo {volume} {92}},\ \bibinfo
  {pages} {052111} (\bibinfo {year} {2015})}\BibitemShut {NoStop}%
\bibitem [{\citenamefont {Nowak}\ and\ \citenamefont
  {Tarnowski}(2017{\natexlab{a}})}]{nowak2017jstat}%
  \BibitemOpen
  \bibfield  {author} {\bibinfo {author} {\bibfnamefont {M.~A.}\ \bibnamefont
  {Nowak}}\ and\ \bibinfo {author} {\bibfnamefont {W.}~\bibnamefont
  {Tarnowski}},\ }\bibfield  {title} {\bibinfo {title} {Spectra of large
  time-lagged correlation matrices from random matrix theory},\ }\href
  {https://doi.org/10.1088/1742-5468/aa6504} {\bibfield  {journal} {\bibinfo
  {journal} {J. Stat. Mech.: Theory Exp.}\ }\textbf {\bibinfo {volume}
  {2017}}\bibinfo  {number} { (6)},\ \bibinfo {pages} {063405}}\BibitemShut
  {NoStop}%
\bibitem [{\citenamefont {Nowak}\ and\ \citenamefont
  {Tarnowski}(2017{\natexlab{b}})}]{nowak2017pre}%
  \BibitemOpen
\bibfield  {number} {  }\bibfield  {author} {\bibinfo {author} {\bibfnamefont
  {M.~A.}\ \bibnamefont {Nowak}}\ and\ \bibinfo {author} {\bibfnamefont
  {W.}~\bibnamefont {Tarnowski}},\ }\bibfield  {title} {\bibinfo {title}
  {Complete diagrammatics of the single-ring theorem},\ }\href
  {https://doi.org/10.1103/PhysRevE.96.042149} {\bibfield  {journal} {\bibinfo
  {journal} {Phys. Rev. E}\ }\textbf {\bibinfo {volume} {96}},\ \bibinfo
  {pages} {042149} (\bibinfo {year} {2017}{\natexlab{b}})}\BibitemShut
  {NoStop}%
\bibitem [{\citenamefont {Jarosz}(2011)}]{jarosz2011}%
  \BibitemOpen
  \bibfield  {author} {\bibinfo {author} {\bibfnamefont {A.}~\bibnamefont
  {Jarosz}},\ }\bibfield  {title} {\bibinfo {title} {Summing free unitary
  random matrices},\ }\href {https://doi.org/10.1103/PhysRevE.84.011146}
  {\bibfield  {journal} {\bibinfo  {journal} {Phys. Rev. E}\ }\textbf {\bibinfo
  {volume} {84}},\ \bibinfo {pages} {011146} (\bibinfo {year}
  {2011})}\BibitemShut {NoStop}%
\end{thebibliography}%

\end{document}